\documentclass[aip,jcp,12pt]{revtex4-2}
\pdfoutput=1

\usepackage{graphicx}% Include figure files
\usepackage{dcolumn}% Align table columns on decimal point
\usepackage{bm}% bold math
\usepackage{tabularx} 
\usepackage[utf8]{inputenc}
\usepackage[T1]{fontenc}
\usepackage{mathptmx}
\usepackage{etoolbox}
\usepackage{booktabs}
\usepackage[table]{xcolor}

\newcolumntype{L}[1]{>{\raggedright\let\newline\\\arraybackslash\hspace{0pt}}m{#1}}
\newcolumntype{C}[1]{>{\centering\let\newline\\\arraybackslash\hspace{0pt}}m{#1}}

\usepackage[font=scriptsize]{caption}
\usepackage{caption}
\usepackage{amsmath}
\usepackage{amssymb}
\usepackage{float}
\usepackage{chemfig}
\usepackage{multirow}
\usepackage{subcaption}
\usepackage{natbib}

\usepackage{pdfpages}
\makeatletter
\AtBeginDocument{\let\LS@rot\@undefined}
\makeatother

\makeatletter
\def\@email#1#2{%
 \endgroup
 \patchcmd{\titleblock@produce}
  {\frontmatter@RRAPformat}
  {\frontmatter@RRAPformat{\produce@RRAP{*#1\href{mailto:#2}{#2}}}\frontmatter@RRAPformat}
  {}{}
}%
\makeatother
\begin{document}

%\preprint{AIP/123-QED}

\title{Theoretical design of nanocatalysts based on (Fe$_2$O$_3$)$_n$ clusters for hydrogen production from ammonia}
% Force line breaks with \\
%% Authors and affiliations:

\author{Sapajan Ibragimov}
\affiliation{Faculty of Chemistry, Gda\'nsk University of Technology, Narutowi za 11/12, 80-233, Gda\'nsk, Poland}%Lines break automatically or can be forced with \\
\affiliation{Faculty of Technical Physics and Applied Mathematics, Gda\'nsk University of Technology, Narutowicza 11/12, 80-233 Gda\'nsk, Poland}

\author{Andrey Lyalin}
\email[Author to whom correspondence should be addressed: ]{lyalin@icredd.hokudai.ac.jp}
\affiliation{Department of Chemistry, Faculty of Science, Hokkaido University, Sapporo 060-0810, Japan}
\affiliation{Research Center for Energy and Environmental Materials (GREEN), National Institute for Materials Science, Namiki 1-1, Tsukuba 305-0044,Japan}%

\author{Sonu Kumar} 
\affiliation{Institute for Chemical Reaction Design and Discovery (WPI-ICReDD), Hokkaido University, Sapporo 001-0021, Japan}

\author{Yuriko Ono} 
\affiliation{Institute for Chemical Reaction Design and Discovery (WPI-ICReDD), Hokkaido University, Sapporo 001-0021, Japan}

\author{Tetsuya Taketsugu}
\affiliation{Department of Chemistry, Faculty of Science, Hokkaido University, Sapporo 060-0810, Japan}
\affiliation{Institute for Chemical Reaction Design and Discovery (WPI-ICReDD), Hokkaido University, Sapporo 001-0021, Japan} 

\author{Maciej Bobrowski}
\affiliation{Faculty of Technical Physics and Applied Mathematics, Gda\'nsk University of Technology, Narutowicza 11/12, 80-233 Gda\'nsk, Poland}

\date{\today}% It is always \today, today,

\begin{abstract}
The catalytic activities of high-spin small Fe(III) oxides have been investigated for efficient hydrogen production through ammonia decomposition, using the Artificial Force Induced Reaction (AFIR) 
method within the framework of density functional theory (DFT) with the B3LYP hybrid exchange-correlation functional. 
Our results reveal that the adsorption free energy of NH$_3$ on (Fe$_2$O$_3$)$_n$ ($n=1-4$) decreases with increasing cluster size up to $n=3$, followed by a slight increase at $n=4$. 
The strongest NH$_3$ adsorption energy, 33.68 kcal/mol, was found for Fe$_2$O$_3$, where NH$_3$ interacts with a two-coordinated Fe site, forming an Fe-N bond with a length of 2.11 \AA. 
A comparative analysis of NH$_3$ decomposition and H$_2$ formation on various Fe(III) oxide sizes identifies the rate-determining steps for each reaction. 
We found that the rate-determining step for the full NH$_3$ decomposition on (Fe$_2$O$_3$)$_n$ ($n=1-4$) is size-dependent, with the NH$^{*}$ $\rightleftharpoons$ N$^{*}$ + 3H$^{*}$ reaction acting as the limiting step for $n=1-3$. 
Additionally, our findings indicate that H$_2$ formation is favored following the partial decomposition of NH$_3$ on Fe(III) oxides.
\end{abstract}

\maketitle

%%%%%%%%INTRODUCTION%%%%%%%%%%%%%%%%%%%%%%%%%%%%%
%%%------------------%%%%%%%%%%%%%%%%%%%%%%%%%%%%
\section{\label{sec:level1}Introduction:\protect\\}

The ammonia decomposition reaction has recently received extensive attention due to its potential use as an alternative green energy source \cite{hansgen2010using,plana2010ni,liu2008promotion,lan2012ammonia,klerke2008ammonia}. 
One of the key advantages of ammonia as a green energy source is its ability to be liquefied at low pressure and a relatively low temperature of 20 $^\circ$C, 
making it an attractive candidate for hydrogen storage and transportation. As with many other chemical processes, catalysts play a crucial role in ammonia decomposition to achieve fast and efficient H$_2$ production. 
Experimental and theoretical studies have demonstrated that Ru-based catalysts are the most active for ammonia decomposition \cite{ganley2004priori,yin2004investigation,logadottir2003ammonia}. 
However, ruthenium's high cost and limited availability pose challenges for its large-scale industrial application. 
Therefore, developing new types of cost-effective catalysts for NH$_3$ decomposition, based on non-noble metals or metal oxides, 
has become a significant area of research for effective hydrogen generation \cite{lucentini2021review}. 
Numerous studies have focused on the activity of catalysts involving various metals and alloys \cite{yao2011core}. 
Among the most studied non-noble metals, iron (Fe) stands out as a leading catalyst due to its low cost and availability. While the reactivity of Fe is lower compared to other transition metals, 
it can be enhanced by using nanoparticles instead of extended surfaces. 
Indeed, it is well known that the reactivity of small-size clusters can be finely tuned by adjusting their size, geometry, and electronic structure, 
making them promising catalysts in various catalytic processes \cite{zemski2002studies,yang2020surface,tyo2015catalysis,heiz2007nanocatalysis,fernando2015quantum}. 
For example, Nishimaki, et al. \cite{nishimaki1999formation} experimentally studied ammonia decomposition on Fe nanoparticles of various grain sizes (20 nm to 1 $\mu$m) in an ammonia steam environment. 
Their findings indicated that the highly reactive surface of nanoparticles enhances NH$_3$ dissociation without increasing the nitrogen content in the gas phase, 
resulting in nitride phases that depend on the grain size and morphology.

As an alternative approach, ammonia decomposition reactions on small nanosized Fe clusters are frequently investigated using density functional theory (DFT) methods. 
Theoretical studies suggest that the mechanisms of ammonia decomposition involve stepwise dehydrogenation, where the rate-limiting step can 
vary depending on the size, type, and shape of the catalysts. Thus, G. Lanzani and K. Laasonen employed spin-polarized DFT to examine the adsorption 
and dissociation of NH$_3$ on a single nanosized icosahedral Fe$_{55}$ cluster \cite{lanzani2010nh3}. Their research indicated that the overall reaction barrier for 
stepwise dehydrogenation was 1.48 eV, with different active sites on the Fe$_{55}$ cluster (facets and vertices), where the rate-limiting step was the initial hydrogen dissociation. 
Similarly, G.S. Otero et al. \cite{otero2016evaluating} conducted a comprehensive comparative study on various sizes of
 Fe clusters (Fe$_{16}$, Fe$_{22}$, Fe$_{32}$, Fe$_{59}$, Fe$_{80}$, Fe$_{113}$, Fe$_{190}$) and Fe(111) surfaces with additional adatoms. 
Their findings indicated that the reaction kinetics were influenced more by the strength of NH$_3$ adsorption rather than the activation energy barrier. 
Stronger NH$_3$ adsorption led to enhanced dissociation compared to desorption. The studies mentioned above primarily focus on the catalytic activities 
of large Fe clusters and Fe surfaces in the ammonia decomposition reaction. 
However, Xilin Zhang et al. \cite{zhang2015adsorption} specifically investigated the activities of relatively small Fe clusters, 
ranging from single Fe atoms to Fe$_4$ clusters. They found that the highest catalytic activity for stepwise NH$_3$ dehydrogenation was observed with nonatomic iron clusters. 
Interestingly, they observed that the rate-limiting steps differed: co-absorbate NH dissociation for Fe and Fe$_3$, and co-absorbate NH$_2$ dissociation for Fe$_2$ and Fe$_4$.

The NH$_3$ decomposition reaction can be enhanced in the presence of oxygen, where it can proceed through various pathways, 
including ammonia oxidation and hydrogen evolution reactions. Moreover, metal oxides are commonly employed as catalyst supports 
in ammonia decomposition to enhance dispersion and catalytic stability. 
Among these supports, widely used materials include Al$_2$O$_3$, TiO$_2$, as well as carbon nanotubes 
and nanofibers \cite{zhang2005characterizations,karim2009correlating,yin2004investigation,ji2013fe,zhang2007commercial,xuezhi2010carbon}. 
However, metal oxides not only serve as support but also play a crucial role in hydrogen evolution reactions in electrocatalysis, 
where the oxidation state of metals significantly influences the catalytic activity of ammonia decomposition. 
In particular, iron-based oxides, such as Fe$_2$O$_3$, are extensively studied forms of iron oxide due to their low cost and abundance, 
although their activity and stability can vary depending on their structure 
and size \cite{yang2016structural,iordanova2005charge,sanchez2017iron,machala2011polymorphous,yingying2016efficient,SHI2022111681,zhao2020electrocatalytic}.

In this work, we elucidate the role of the size- and structural effects on the catalytic activity of iron-oxide-based nano-catalysts 
toward efficient ammonia decomposition. In particular, we investigated the theoretical mechanisms of stepwise 
ammonia decomposition on (Fe$_2$O$_3$)$_n$ clusters with $n=1-4$ to compare the reactivity of different-sized Fe(III) oxides 
using the Artificial Force Induced Reaction (AFIR) method \cite{maeda2014exploring,sameera2016artificial}.
Additionally, we examined the NH$_3$ adsorption and various energy barriers for NH$_3$ dehydrogenation on different active sites of Fe(III) oxides. 
Our investigation aims to contribute to the design of nanocatalysts based on Fe$_2$O$_3$ by exploring the activity of small-sized Fe(III) oxide clusters.

%==================================================
%=========Computational details====================
%==================================================
\section{Computational details}

All calculations were performed using spin-unrestricted Kohn-Sham DFT with Becke's three-parameter hybrid functional combined with the Lee, Yang, and Parr correlation functional, 
denoted as B3LYP \cite{1B3LYP,2B3LYP,3B3LYP}. In our calculations we have employed the LANL2DZ \cite{1LANL2DZ,2LANL2DZ,3LANL2DZ} 
basis set with effective core potentials (ECP), as well as the Pople-style 6-31+G* basis set, equivalent to 6-31+G(d), 
which includes polarization (d) and diffuse (sp) functions, as it is implemented in the Gaussian 16 program \cite{g16}. 
These methods have been successfully applied to metals and metal oxide systems in previous studies. 
Thus, Glukhovtsev et al. \cite{glukhovtsev1997performance} reported that the performance of the B3LYP/ECP method for systems 
containing iron with various types of bonding showed good agreement with experimental data and high-level theoretical methods (CCSD(T), MCPE, CASSCF). 
Similarly, Taguchi, et al. \cite{taguchi2008new} studied Fe$_6$O$_2$(NO$_3$)$_4$(hmp)$_8$(H$_2$O)$_2$$_2$, [Fe$_4$(N$_3$)$_6$(hmp)$_6$], and Fe$_8$O$_3$(OMe)(pdm)$_4$(pdmH)$_4$(MeOH)$_2$$_5$ 
clusters using the B3LYP/LANL2DZ level of theory, obtaining results that were consistent with experimental data.

At the initial stage, the most stable isomers of iron trioxide for each selected size were investigated using the DFT method. 
A single iron trioxide molecule contains two Fe$^{3+}$ ions; therefore, there are often several energetically 
accessible spin states (0, 1, 2, 3, 4, 5). For the starting cluster Fe$_2$O$_3$, the lowest energy structure corresponds 
to the nonet state with a total spin $S$=4. For (Fe$_2$O$_3$)$_2$, the lowest energy solution was found with a 
total spin $S$=10, indicating an increase in the number of Fe$^{3+}$ ions, which raises the total spin projection. 
For (Fe$_2$O$_3$)$_3$, the lowest energy structure was found with a total spin $S$=15, and lastly, in the case of (Fe$_2$O$_3$)$_4$, 
the lowest energy structure had a total spin $S$=20. Therefore, all clusters considered in our study were in a ferromagnetic configuration. 
We confirmed that spin contamination in the low-lying energy structures was negligible.

To analyze the most favorable pathways of NH$_3$ dehydrogenation and H$_2$ formation reactions catalyzed by small (Fe$_2$O$_3$)$_n$ ($n=1-4$) clusters, 
we applied the SC-AFIR and DS-AFIR methods implemented in the Global Reaction Route Mapping (GRRM) 
strategy \cite{maeda2010communications,maeda2011finding,maeda2013systematic,maeda2014exploring,maeda2018implementation}. 
These automated reaction path search methods have been successfully applied to many catalytic reactions in combination with 
DFT methods \cite{sharma2017dft,skjelstad2022oxyl,sameera2016artificial,gao2014application,gao2015reactivity}. 
The basic idea in the AFIR strategy is to push fragments (reactants) A and B of the whole system together or pull them apart by minimizing the following AFIR function \cite{maeda2014exploring}:
\begin{equation} 
\label{eq:afir}
F(Q) = E(Q) + \alpha\frac{\sum_{i \in A}^{}\sum_{i \in B}^{}\omega_{ij}r_{ij}}{\sum_{i \in A}^{}\sum_{i \in B}^{}\omega_{ij}}
\end{equation} 

The external force term in (\ref{eq:afir}) perturbs the given adiabatic Potential Energy Surface (PES), $E(Q)$, with geometrical parameters ${Q}$ in the AFIR function. 
Here, $\alpha$ defines the strength of the artificial force which depends on the weighted sum of the interatomic distances $r_{ij}$ between atoms $i$ and $j$, with 
the weighths $\omega_{ij}$ defined as 
 \begin{equation}
\label{eq:omega}
\omega_{ij} = \left[ \frac{R_i+R_j}{r_{ij}} \right]^6,
\end{equation}
\noindent where ${R_i}$ and ${R_j}$ the covalent radii of of atoms ${i}$ and ${j}$, respectively.  
The force parameter $\alpha$ in (\ref{eq:afir}) can be expressed as follows:
\begin{equation} \label{eq:alpha}
\alpha = \frac{\gamma}{ \left[ 2^{-1/6} - (1 + \sqrt{1 + \gamma/\epsilon})^{-{1}/{6}} \right] R_{0} },
\end{equation}  
\noindent where $R_0$ and $\epsilon$ are parameters corresponding to interatomic Lennard-Jones potentials, and parameter $\gamma$ has a physical meaning of a collision energy.

This perturbation of the PES facilitates the exploration of additional approximate transition states and local minima on the surface. 
The model collision energy parameter $\gamma$ in (\ref{eq:alpha}) serves as an approximate upper limit for the barrier height that the system 
can be affected by the AFIR function \cite{maeda2014exploring}. In our calculations, $\gamma$ was set to 300 kJ/mol for the entire system. 
During the initial reaction path search, the LANL2DZ basis set was applied with an artificial force to yield approximate products and 
transition states (TS). Subsequently, we utilized the 6-31+G* basis set to optimize these approximate transition states and local minima 
without the artificial force, employing the Locally Updated Planes (LUP) method. 
The vibrational frequency calculations have been performed to confirm the nature of the stationary points, whether they are minima or transition states. 
The results presented in this paper include reaction route mapping at the B3LYP/LANL2DZ level and reaction pathways at the B3LYP/6-31+G(d) level.

The bindig energy $E_{\rm b}$ per unit $n$ of a (Fe$_2$O$_3$)$_n$ cluster is defined as follows: 
\begin{equation}
\centering
E_{\rm b} = - \frac{E_{\rm el}(({\rm Fe}_{2}{\rm O}_{3})_{n}) + E_{\rm ZPE}(({\rm Fe}_{2}{\rm O}_{3})_{n})) - [2n E({\rm Fe})+ 3n E({\rm O})]}{n}
\label{eq:E_binding}
\end{equation}
\noindent where \textit{E$_{\rm el}(({\rm Fe}_{2}{\rm O}_{3})_{n})$} and \textit{E$_{\rm ZPE}(({\rm Fe}_{2}{\rm O}_{3})_{n})$} are the electronic and zero-point energies 
of a cluster (Fe$_2$O$_3$)$_n$ with a number of units $n$, 
while $E$(Fe) and $E$(O) are the energis of free Fe and O atoms.

The standard free energy of adsorption, $\Delta G_{ads}$, is given as 
\begin{equation}
\centering 
\Delta G_{ads} = G{({\rm NH}_{3}@({\rm Fe}_2{\rm O}_3)_n)} - (G({({\rm Fe}_2{\rm O}_3)_n}) + G({{\rm NH}_3}) ) 
\label{eq:ads}
\end{equation}

\noindent where \textit{G${({\rm NH}_{3}@({\rm Fe}_2{\rm O}_3)_n)}$} is the free energy of the most stable structure of the (Fe$_2$O$_3$)$n$ cluster with the adsorbed ammonia molecule, 
\textit{G${({\rm Fe}_2{\rm O}_3)_n}$)} is the free energy of the bare (Fe$_2$O$_3$)$n$ cluster, and \textit{G(${{\rm NH}_3}$)} is the free energy of a single ammonia molecule.
The values of free energy \textit{G} in (\ref{eq:ads}) can be calculated as follows:

\begin{equation}
\centering 
G = E_{\rm el} + E_{\rm ZPE} - TS, 
\label{eq:ads_P}
\end{equation}

\noindent where \textit{E$_{\rm el}$} and \textit{E$_{\rm ZPE}$} are the electronic and zero-point energies of the system, \textit{S} is the entropy of the system, and \textit{T} is the temperature.

%====================================================
%==============Reaction mechanisms============================
%====================================================
\section{Results and discussion}
\label{sec:obj}

In the present work we systematically investigated the ammonia decomposition reaction mechanisms on (Fe$_2$O$_3$)$_n$ clusters of various sizes $n$, where $n=1-4$.
Firstly, we identified approximate reaction pathways for the interactions between NH$_3$ molecules and the most stable isomers of (Fe$_2$O$_3$)$_n$ clusters using the AFIR technique. 
The obtained AFIR pathways were subsequently re-optimized along the minimum energy path using the Locally Updated Plane (LUP) method, without applying artificial forces.
We calculated various reaction mechanisms and the stepwise dissociation\cite{hellman2009ammonia} of hydrogen atoms from nitrogen-containing compounds on Fe(III) oxide clusters, 
following the elementary steps:
\begin{equation}
\centering
{\rm NH}_{3} + ^{*} \rightleftharpoons {\rm NH}^{*}_{3}  
\label{eq:first}
\end{equation}

\begin{equation}
\centering
{\rm NH}^{*}_{3}  \rightleftharpoons {\rm NH}^{*}_{2} + {\rm H}^{*} 
\label{eq:second}
\end{equation}

\begin{equation}
\centering
{\rm NH}^{*}_{2} \rightleftharpoons {\rm NH}^{*} + 2{\rm H}^{*}
\label{eq:third}
\end{equation}

\begin{equation}
\centering
{\rm NH}^{*}  \rightleftharpoons {\rm N}^{*} + 3{\rm H}^{*},
\label{eq:fourth}
\end{equation}

Here $^{*}$ denotes the adsorbed intermedeates on the (Fe$_2$O$_3$)$_n$ cluster's surface. 
Finally, the adsorbed hydrogen atoms on the (Fe$_2$O$_3$)$_n$ clusters can combine to produce molecular hydrogen (H$_2$):

\begin{equation}
\centering
{\rm NH}^{*} + 2{\rm H}^{*} \rightleftharpoons {\rm NH}^{*} + {\rm H}_{2}
\label{eq:h2_one}
\end{equation}

\begin{equation}
\centering
{\rm N}^{*} + 3{\rm H}^{*} \rightleftharpoons {\rm N}^{*} + {\rm H}^{*} + {\rm H}_{2}
\label{eq:last}
\end{equation}
 
The paper is organized as follows. We first discuss the structures of free clusters, followed by the adsorption 
of NH$_3$ on the most stable isomers of (Fe$_2$O$_3$)$_n$, $n=1-4$, clusters. 
We then examine the complete dehydrogenation and H$_2$ formation processes for each cluster size.

\subsection{\textbf{Structure of (Fe$_2$O$_3$)$_n$ clusters with $n=1-4$}} 

\begin{figure}[H]
\centering
\includegraphics[width=12cm]{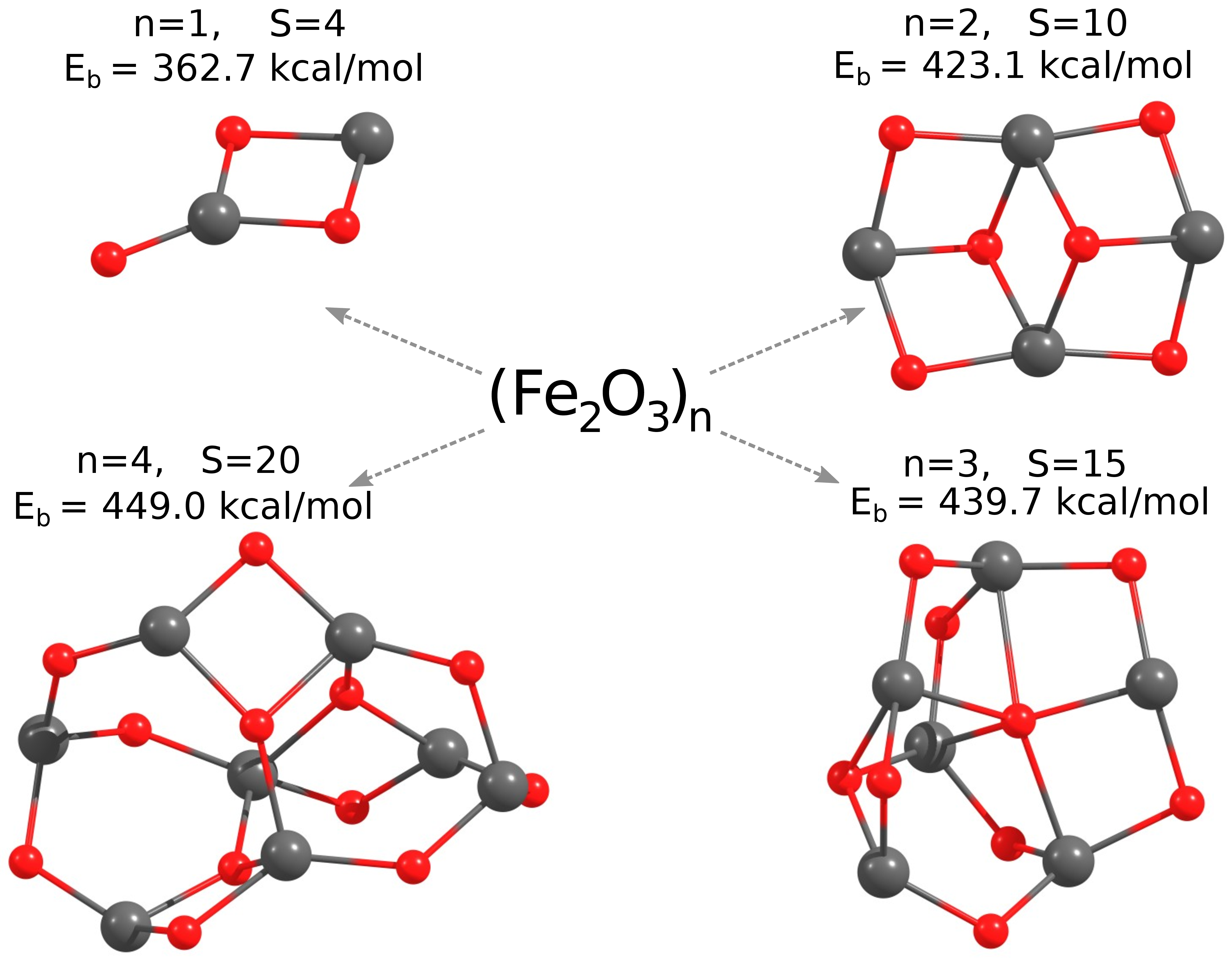}
\caption{The most stable structures of (Fe$_2$O$_3$)$_n$ clusters with $n=1-4$. The total spin $S$  and the binding energy $E_{\rm b}$, of the clusters are shown in inserts.
} 
\label{fig:Fe2O3_n}
\end{figure}

{Figure~\ref{fig:Fe2O3_n}} demonstrates the most stable structures of small (Fe$_2$O$_3$)$_n$ clusters with $n=1-4$, as obtained in the present work using automated GRRM approach.
A total of up to 60 isomer structures have been obtained for each cluster size n. The low-energy isomers for each cluster size, along with their relative binding energies are presented in Figs. S1-S4. 
The most stable structures are consistent with those obtained in our previous DFT study, which employed {B3LYP functional} and 
five different types of basis sets (LANL2DZ, 6-31+G$^{*}$, 6-311+G$^{*}$, Sapporo-DZP, and aug-cc-pVTZ).{\cite{Sapajan2024}}
We found that the most stable structure of the smallest Fe$_2$O$_3$ cluster is a nonet kite-like type with a binding energy $E_{\rm b}$=362.7 kcal/mol. 
The kite-like structure is a commonly studied configuration \cite{shiroishi2003structure,jones2005structural} and was previously investigated by Sierka et al. \cite{erlebach2014structure}, 
who observed the most stable spin configuration for this structure to be {$S$=0}. 
In contrast, we found that the lowest energy structure corresponds to a nonet state with $S$=4, 
while the singlet kite-like structure is 0.62 kcal/mol less stable {as shown in Table S1}. 
The results of our calculations show that the absolute binding energy of (Fe$_2$O$_3$)$_n$ rapidly increases with increasing cluster size $n$ from 1 to 2 by 60.4 kcal/mol. 
However, further growth in binding energy with cluster size slows down, demonstrating a tendency for saturation as $n$ increases.

\subsection{\textbf{Ammonia adsorption on (Fe$_2$O$_3$)$_n$ clusters}}
\label{sec:adsorption}

\begin{figure}[H]
\centering
\includegraphics[width=12cm]{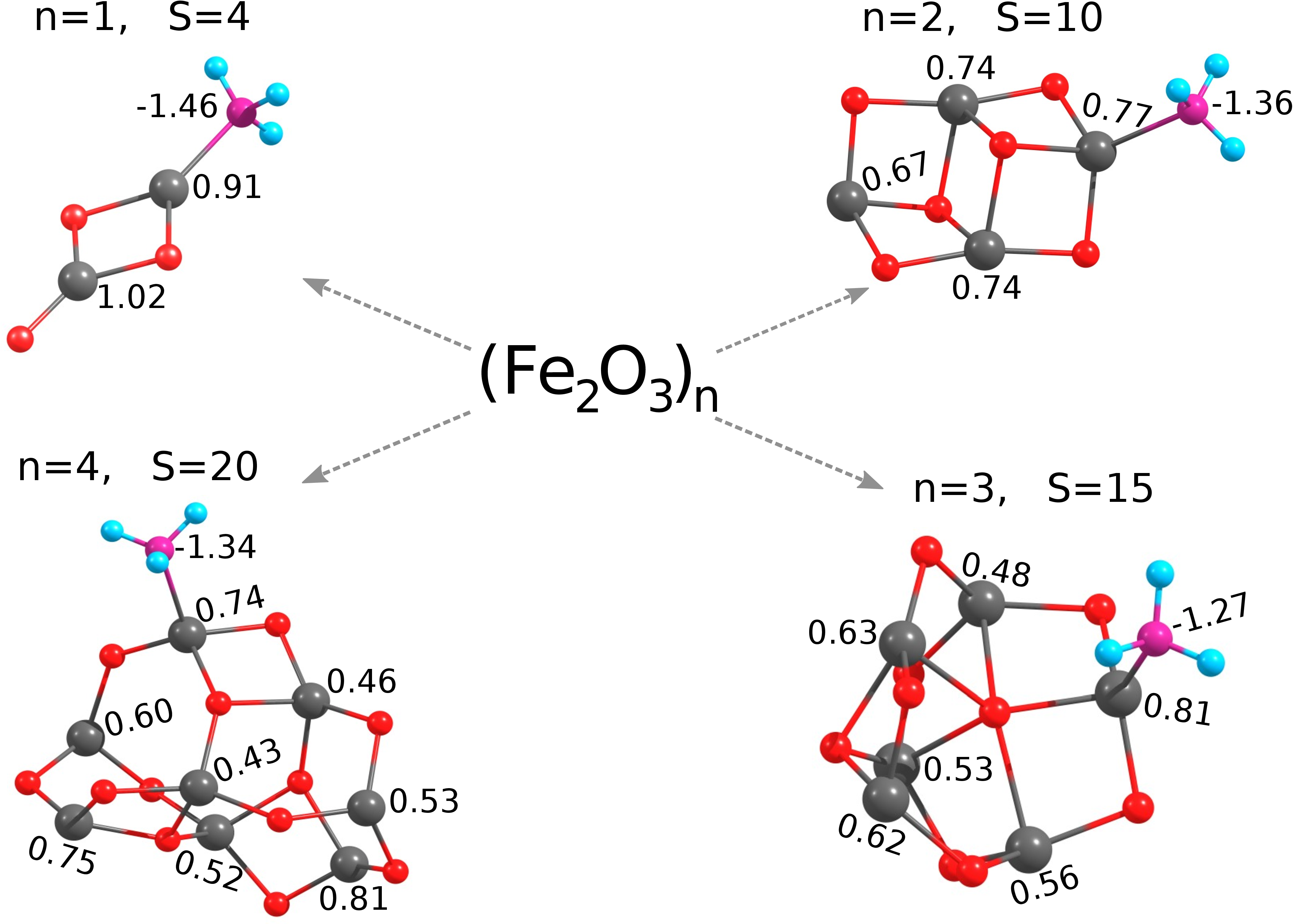}
\caption{The optimized geometries of NH$_3$ on (Fe$_2$O$_3$)$_n$ for $n = 1-4$; N$-$Fe distances (\AA) are shown in parentheses along with the partial atomic charges on 
neighboring atoms. The total spin $S$ of the clusters is shown in inserts.}
\label{fig:adsorption_n}
\end{figure}

Adsorption of ammonia on (Fe$_2$O$_3$)$_n$ clusters is a crucial initial step in the whole dehydrogenation process.
Figure~\ref{fig:adsorption_n} demonstrates the most stable adsorption configurations of the NH$_3$ molecule on (Fe$_2$O$_3$)$_n$ clusters with $n=1-4$.
The corresponding free energies of adsorption and Fe$-$N bond distances are shown in Table~\ref{tab:all_n} {at 0 K}. 
Our calculations show that the adsorption of NH$_3$ on the smallest Fe$_2$O$_3$ cluster is the most stable among all cluster sizes considered in this study,
with an adsorption free energy of {-33.68 kcal/mol}. This finding is corroborated by Mulliken charge analysis, which shows that more electrons are shared between
the lone pair of the N atom and the 3d orbitals of Fe$^{2+}$ for $n=1$.
On the other hand, for larger cluster sizes with $n=2-4$, which primarily contain Fe$^{3+}$, the electron density is more localized over the bonding region,
as also reported by Sierka et al. \cite{erlebach2014structure}. Therefore, bonding occurs with the nitrogen lone pair.

Our theoretical analysis indicates that the adsorption energy $\Delta$G$_{ads}$ of ammonia on (Fe$_2$O$_3$)$_n$ clusters decreases from $n=1$ to $n=3$,
followed by a slight increase for $n=4$. A similar trend in the change of adsorption energy with cluster size was reported by Shulan Zhou et al.\cite{zhou2018first} for Ru$_n$@CNT systems.
We also compared the adsorption energy of NH$_3$ on different metal and metal oxides in Table~\ref{tab:all_n}.
The obtained NH$_3$ adsorption energies on (Fe$_2$O$_3$)$_n$ clusters are about 10 kcal/mol higher than the data reported by Zhang et al. for the Ru(0001) surface \cite{zhang2002density}.
Moreover, the adsorption of NH$_3$ and NO$_x$ on the $\gamma$-Fe$_2$O$_3$(111) surface was studied by Wei Huang et al. \cite{huang2023density}
using periodic density functional calculations. They calculated adsorption energies on octahedral and tetrahedral sites of $\gamma$-Fe$_2$O$_3$(111)
to be { -2.13 kcal/mol} and -21.68 kcal/mol, respectively.
Similarly, our calculated NH$_3$ adsorption energies on (Fe$_2$O$_3$)$_n$ clusters for $n=3$ and $n=4$ are close to the data reported by Wei Huang et al.,\cite{huang2023density}
as adsorption of NH$_3$ on the three-coordinated Fe$^{3+}$ site resembles the tetrahedral site of $\gamma$-Fe$_2$O$_3$(111),
while the adsorption on the four-coordinated Fe$^{3+}$ site resembles the octahedral site of $\gamma$-Fe$_2$O$_3$. 

\begin{table}[H]
\centering
\caption{\label{tab:all_n} NH$_3$ adsorption free energy $\Delta$G$_{ads}$ and d(Fe -- N) bond length in various size of (Fe$_2$O$_3$)$_n$ where $n=1-4$.}
\begin{tabular}{llllllllll} 
       & & $\Delta$G$_{ads}$ \textit{(kcal/mol)}  & {Fe $-$ N} (\AA)  & & reference      \\ \hline \hline
NH$_3$/(Fe$_2$O$_3$)$_1$           & & -33.68    & 2.11  &   &   \\
NH$_3$/(Fe$_2$O$_3$)$_2$           & & -30.97    & 2.14  &   & this work  \\ 
NH$_3$/(Fe$_2$O$_3$)$_3$           & & -30.36    & 2.15  &   &   \\
NH$_3$/(Fe$_2$O$_3$)$_4$           & & -30.59    & 2.14  &   &    \\
\hline
\multirow{2}{*}{NH$_3$/ZnFe$_2$O$_4$(110)}    & &    -48.54   &  {Zn $-$ N} (2.03)  & &  \multirow{2}{*}{a}   \\
\cline{3-4}
                             & &    -41.52   &  {Fe $-$ N} (1.99)  &  &    \\
\hline
NH$_3$/Ru(0001)              & &    -20.52   &  {Ru $-$ N}  (2.17)   &  & b \\
\hline
\multirow{2}{*}{NH$_3$/Fe$_2$O$_3$/AC}        & &    -49.12, -37.35 &    -      &  &\multirow{2}{*}{c} \\  
\cline{3-4}
                                              & &    -26.29, -31.13 &    -      &   & \\
\hline
NH$_3$/$\gamma$-Fe$_2$O$_3$ nano  & &    -37.52         &    -      &  & d   \\
\hline 
\multirow{2}{*}{NH$_3$/$\gamma$-Fe$_2$O$_3$ (111)} & &    -21.68  & {Fe$_{tet}$ $-$ N} (2.13) & & \multirow{2}{*}{f} \\
\cline{3-4}
                            & &                           -2.13  & {Fe$_{oct}$ $-$ N} (2.101) &  &  \\  
\hline   
a\cite{zou2018nh3}, b\cite{zhang2002density}, c\cite{xie2021adsorption}, d\cite{ren2019study}, f\cite{huang2023density} \\ 
\end{tabular}
\end{table}

%%%%%%%%%%%%%%%%%%%%%%%%%%%%%%%%%%%%%%%%%%%%%%%%%%%%%%%
\iffalse

\begin{figure}[H]
\centering 
\includegraphics[width=12cm]{figures/delta_G_NH3_n_1234_in_K.eps}
\caption{The temperature dependence of adsorption free energy for NH$_3$ adsorption on (Fe$_2$O$_3$)$_n$ $n=1-4$ at 1 atm.}
\label{fig:adsorption_vs_tem}
\end{figure}

\fi
%%%%%%%%%%%%%%%%%%%%%%%%%%%%%%%%%%%%%%%%%%%%%%%%%%%%%%%%%

As mentioned above, the calculated adsorption energies indicate that the adsorption of an NH$_3$ molecule on (Fe$_2$O$_3$)$_n$ clusters ($n = 1-4$) weakens as the cluster size increases from $n = 1$ to $n = 3$. 
In industrial processes, the dehydrogenation of ammonia typically occurs at high temperatures, often in the range of 400$^\circ$C to 700$^\circ$C, depending on the specific catalysts and conditions used.
Therefore, it is important to determine the range of temperatures at which ammonia adsorption on (Fe$_2$O$_3$)$_n$ remains stable.
Figure S5 demonstrates the temperature dependence of $\Delta G_{ads}$ in the range from {0 K} to {1200 K} for the most stable adsorption configurations
of NH$_3$ on (Fe$_2$O$_3$)$n$ clusters ($n = 1-4$). The negative values of $\Delta G_{ads}$ correspond to stable adsorption.
As seen in Fig. S5, NH$_3$ adsorbed on the smallest Fe$_2$O$_3$ cluster is stable 
across the entire temperature range of {0 K} to {1200 K}.
However, for larger cluster sizes, ammonia adsorption becomes energetically unfavorable at 
temperatures of {1107 (K), 961 (K), and 1000 (K)} for $n = 2, 3$, and $4$, respectively.

\subsection{\textbf{NH$_3$ decomposition on Fe$_2$O$_3$}}
\label{sec:n1_NH3}

In this section, we discuss the complete NH$_3$ decomposition and H$_2$ formation reactions { (\ref{eq:first}) - (\ref{eq:last})} on the smallest considered cluster, Fe$_2$O$_3$, at room temperature, T=298.15 K, 
explored by the AFIR method. 
This method allows for the automatic exploration of the full reaction path network, systematically accounting for the variety of possible isomer structures and adsorption sites. 
This is an important approach in nanocatalysis because it has been demonstrated that the most stable structures are not always the most reactive. 
Therefore, a systematic search for reaction pathways that accounts for the contributions of low-energy isomers is required to accurately describe 
the catalytic properties of clusters at finite temperatures.\cite{gao2014application}

To illustrate the isomer and reaction-site effects, we explicitly consider two different isomers of the Fe$_2$O$_3$ cluster: 
the most stable kite-like structure with one terminal oxygen atom, and the linear structure isomer with two terminal oxygen atoms which is 
{6.24 kcal/mol less stable (see Fig.~S1)}.
The kite-like structure possess two type of catalytically active metal centers - two-coordinated and three-coordinated Fe sites. Therefore we consider  
adsorption and decomposition of NH$_3$ molecule on both of them.

\begin{figure}[H]
%\noindent \begin{minipage}[r]{0.45\textwidth}
\centering
\includegraphics[width=12cm]{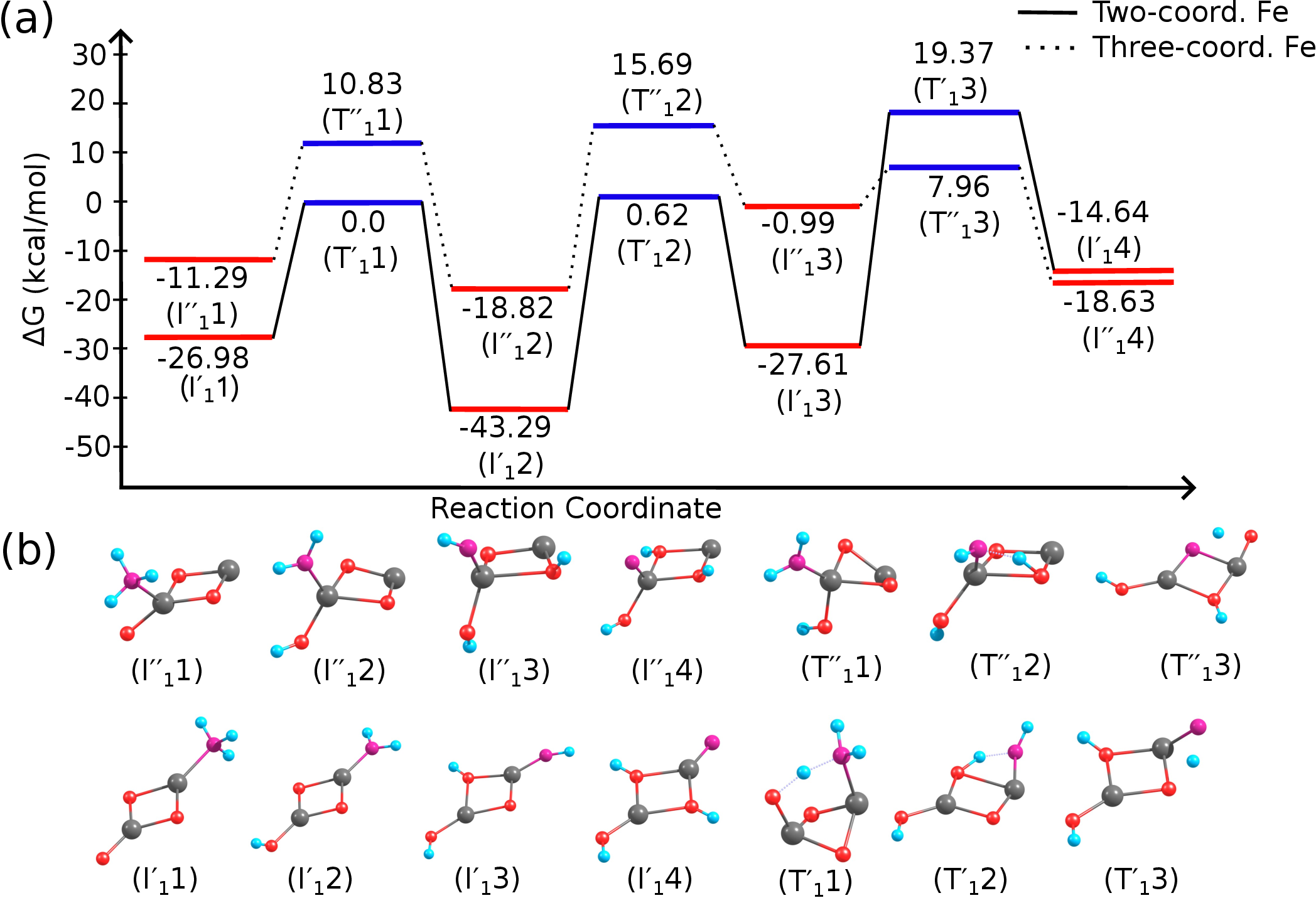}
\caption{(a) The energy profile for NH$^{*}_3$ $\rightarrow$ NH$^{*}_2$ + H$^{*}$ $\rightarrow$ NH$^{*}$ + 2H$^{*}$  $\rightarrow$ N$^{*}$ + 3H$^{*}$  reaction path on the kite-like isomer of 
Fe$_2$O$_3$ at T=298.15 K. (b) Geometries of the optimized equilibrium and transition states along the reaction path.}
\label{fig:case_1}
\end{figure}

\begin{figure}[H]
\centering 
\includegraphics[width=12cm]{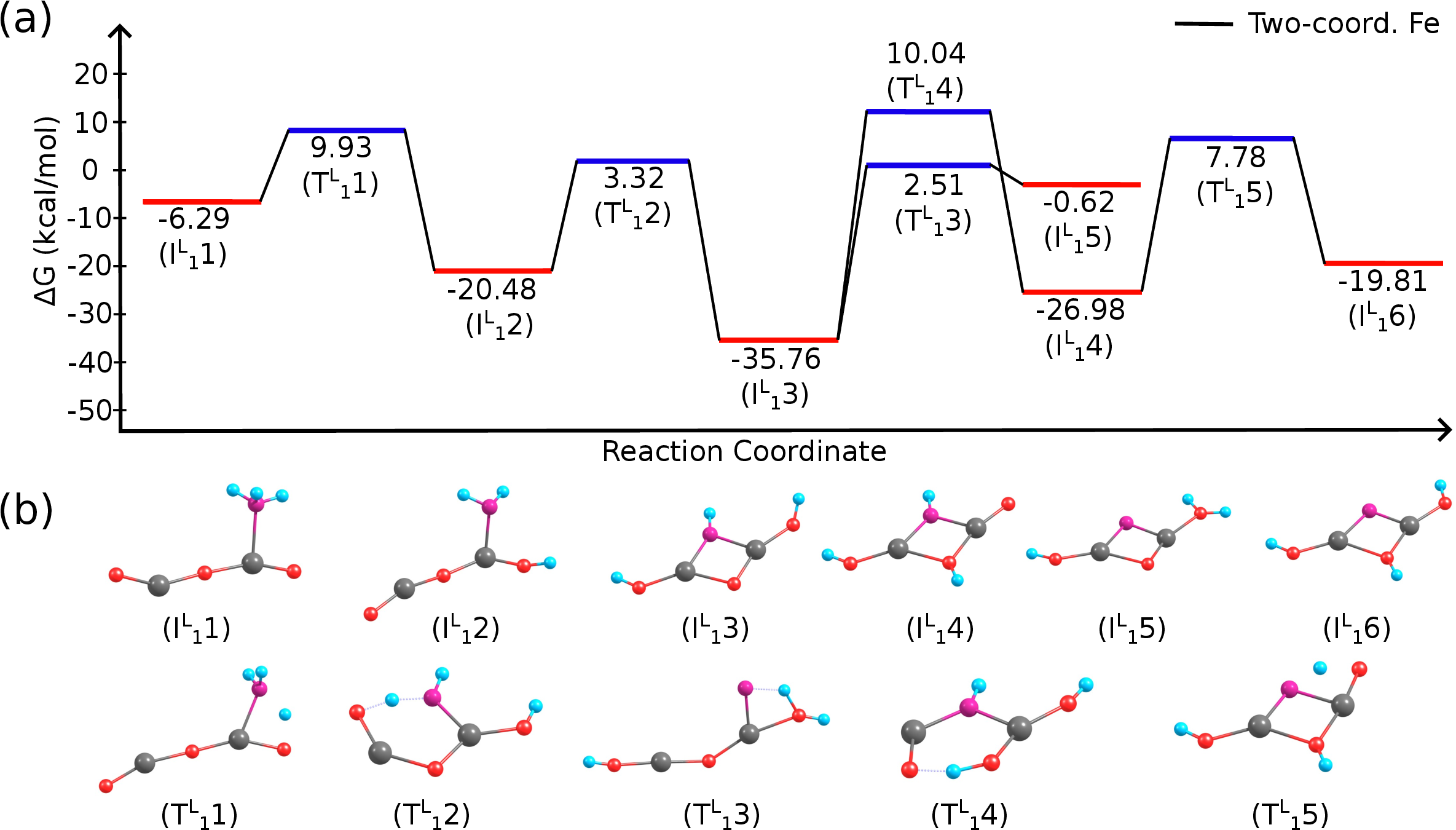}
\caption{(a) The energy profile for NH$^{*}_3$ $\rightarrow$ NH$^{*}_2$ + H$^{*}$ $\rightarrow$ NH$^{*}$ + 2H$^{*}$ + $\rightarrow$ N$^{*}$ + 3H$^{*}$  reaction path on the linear-type isomet of Fe$_2$O$_3$ at T=298.15 K.
(b) Geometries of the optimized equilibrium and transition states along the reaction path.}
\label{fig:case_2}
\end{figure}

{Figure~\ref{fig:case_1}(a)} demonstrates that the adsorption of NH$_3$ on the kite-like Fe$_2$O$_3$ cluster is an exothermic reaction, occurring at both the two-coordinated and three-coordinated Fe sites. 
The adsorption free energies are -26.98 kcal/mol for the two-coordinated Fe site (intermediate I$^{\prime}_{1}$1) and -11.29 kcal/mol for the three-coordinated Fe site 
(intermediate I$^{\prime\prime}_{1}$1), respectively. The optimized structures of all intermediates (I) and transition satates (T) along the reaction pathways 
are shown in Fig.~\ref{fig:case_1}(b) and ~\ref{fig:case_2}(b), for the kite-like and linear clusters, respectively. 
Here the lower index corresponds to the cluster size $n$, while the numbering corresponds to the order of intermedeates (transition states) along the reaction path.
As discussed in the previous section, the most stable adsorption site for NH$_3$ is the two-coordinated Fe site, with an Fe$-$N bond length of 2.11 \AA. 
In contrast, the Fe$-$N bond length at the three-coordinated Fe site is 2.16 \AA. 
These findings are supported by the fact that NH$_3$ adsorption highly depends on the local geometry and electronic structure of the catalyst.

In the case of the Fe$_2$O$_3$ kite-like structure, the first dehydrogenation reaction is the second step in the reaction mechanism, occurring after adsorption with activation 
barriers of 26.98 kcal/mol and 22.12 kcal/mol through the reaction paths I$^{\prime}_{1}$1-T$_{1}^{\prime}$1-I$_{1}^{\prime}$2 and 
I$^{\prime\prime}_{1}$1-T$^{\prime\prime}_{1}$1-I$^{\prime\prime}_{1}$2, respectively. 
The reactions on these two-coodrinated and three-coordinated active sites are exothermic by 16.31 kcal/mol and 7.53 kcal/mol, respectively.
However, the first dehydrogenation of NH$_3$ on the linear-type structure Fig.~\ref{fig:case_2}(a) occurs with smaller activation barrier of 16.22 kcal/mol via the reaction 
path I$^{L}_{1}$1 - T$^{L}_{1}$1 - I$^{L}_{1}$2, demonstrating that the less stable linear isomer is more reactive.

The role of Fe$_2$O$_3$ isomer structure on NH$_3$ adsorption and first hydrogen atom transfer was previousely studied by Chaoyue Xie et al.\cite{xie2021adsorption} 
They performed DFT-D3 calculations on the adsorption 
mechanisms of different molecules (NH$_3$, NO, O$_2$) on activated carbon (AC) supported iron-based catalysts Fe$_x$O$_y$/AC. 
The calculated adsorption electronic energies of NH$_3$ were 
-37.4 kcal/mol and -53.7 kcal/mol  on different isomers of Fe$_2$O$_3$/AC, 
and the first hydrogen atom transfer had an activation barrier of 15.5 kcal/mol. 
Similarly, the adsorption and dehydrogenation of ammonia on different metal oxides were investigated by 
Erdtman and co-workers\cite{erdtman2017simulations} for the application of gas sensors. 
They reported that the adsorption energy of NH$_3$ on the RuO$_2$(110) surface is -38.24 kcal/mol, 
and the first N$-$H bond cleavage had an activation energy barrier of 17.45 kcal/mol.

The third step of the NH$_3$ dehydrogenation reaction (\ref{eq:third}) involves the dissociation of the adsorbed NH$^{*}_2$ intermediate into NH$^{*}$ and H$^{*}$ species. 
In this step, the abstracted hydrogen atom transfers to one of the oxygen atoms in the cluster. 
{Figure~\ref{fig:case_1}(a)} demonstrates, that in the case of the kite-like structure the energy barriers for this step are 43.91 kcal/mol and 34.51 kcal/mol, 
corresponding to the reaction paths I$^{\prime}_{1}$2 - T$^{\prime}_{1}$2 - I$^{\prime}_{1}$3 and I$^{\prime\prime}_{1}$2 - T$^{\prime\prime}_{1}$2 - I$^{\prime\prime}_{1}$3, respectively.

In the fourth step (\ref{eq:fourth}), the adsorbed NH$^{*}$ intermediate further dissociates into N$^{*}$ and H$^{*}$ species as shown in Fig.~\ref{fig:case_1}(a). 
The reaction barriers associated with this step are 46.98 kcal/mol and 8.95 kcal/mol for the two-coordinated and three-coordinated reaction paths, respectively. 
The decomposition of NH$_3$ on kite-like structures becomes endothermic starting from the third step (\ref{eq:third}). 
Our calculations reveal that NH$_3$ dehydrogenation has a high energy barrier when the NH$_3$ molecule is adsorbed at a two-coordinated Fe site, which is the most stable adsorption site.
On the other hand, dehydrogenation of the adsorbed NH$_3$ at a three-coordinated Fe site has a considerably lower activation barrier of 8.95 kcal/mol for the  reaction step (\ref{eq:fourth}).

Overall, for the NH$_3$ decomposition reaction on the kite-like Fe$_2$O$_3$ structure, with initial NH$_3$ adsorption on the two-coordinated Fe atom, 
the rate-limiting step is the fourth reaction (\ref{eq:fourth}), with a barrier of 46.98 kcal/mol. 
Alternatively, for the less favorable NH$_3$ adsorption on the three-coordinated Fe atom, the rate-limiting step is the third reaction step (\ref{eq:third}), with a barrier of 34.51 kcal/mol.

The reaction pathway calculated for NH$_3$ decomposition on the linear-type Fe$_2$O$_3$ isomer is shown in Fig.~\ref{fig:case_2}(a), 
and respective intermediate and transition state structures 
are shown in Fig.~\ref{fig:case_2}(b).   
Since this structure consists of two iron atoms connected through a central oxygen, each containing a terminal oxygen, 
the reaction mechanism differs slightly from that of the kite-like isomer. 
For instance, in the third step of the reaction, the second hydrogen from the adsorbed NH$^{*}_2$ intermediate is transferred to the second terminal oxygen.
The energy barrier for this step on the linear-type structure is 23.8 kcal/mol, as shown in the reaction path (I$^{L}_{1}$2 - T$^{L}_{1}$2 - I$^{L}_{1}$3) in Fig.~\ref{fig:case_2}a.

The fourth step on this isomer is not straightforward, involving the central oxygen atom breaking its bond with one of the neighboring iron atoms while forming an {Fe $-$ N $-$ Fe} bridge.
This process leads to two different intermediates: the formation of the adsorbed H$_2$O$^{*}$ and the transfer of a hydrogen atom from one side of the {Fe $-$ N $-$ Fe} bridge to the other. 
Subsequently, the final dehydrogenation step from the NH$^{*}$ intermediate occurs, with an activation energy barrier of 34.76 kcal/mol.

\begin{figure}[H]
%\noindent \begin{minipage}[r]{0.45\textwidth}
\centering
\includegraphics[width=12cm]{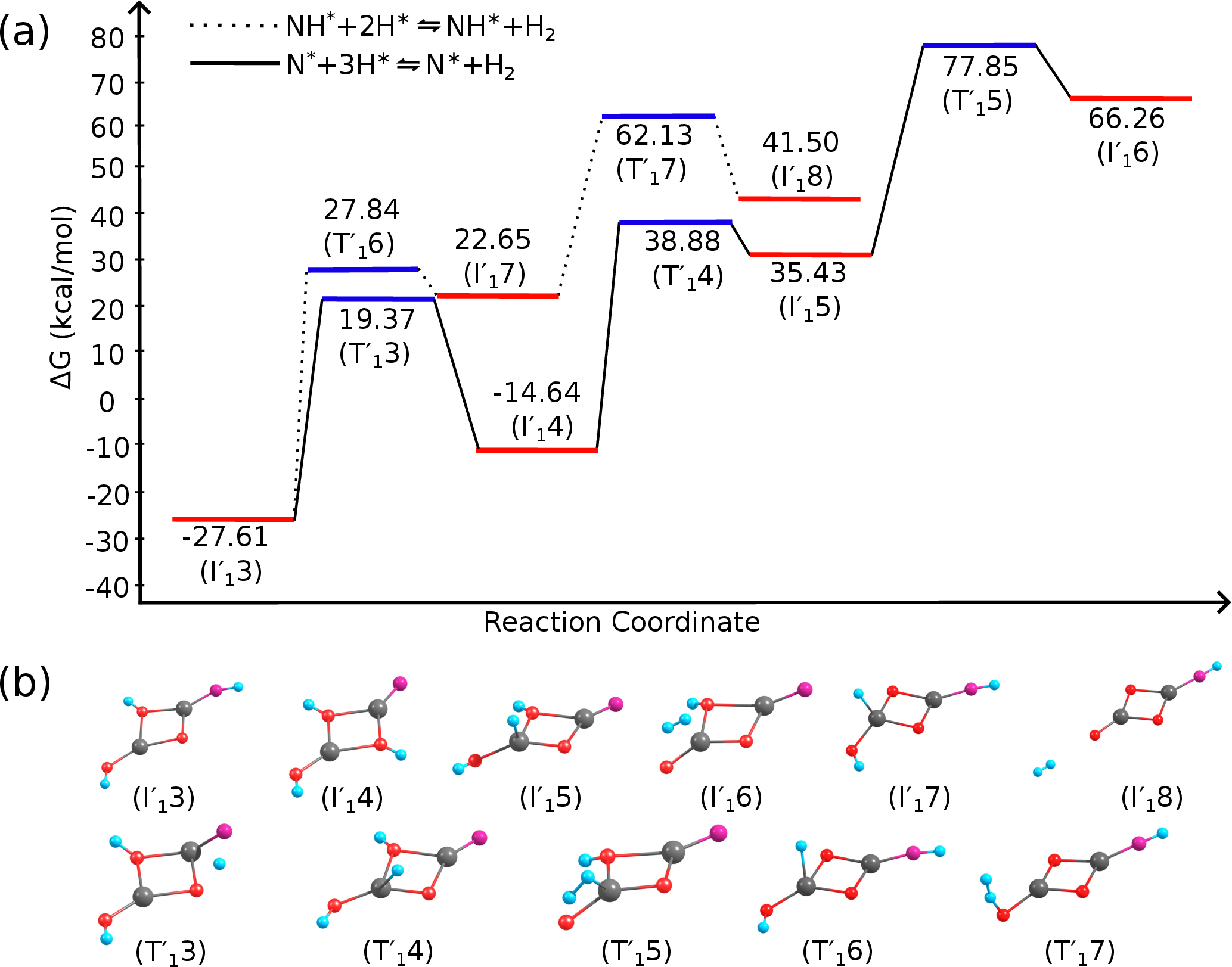}
\caption{(a) The energy profile for H$_2$ formation on the kite-like Fe$_2$O$_3$ cluster at T=298.15 K.
(b) Geometries of the optimized equilibrium and transition states along the reaction path.}
\label{fig:case_1_h2_formation}
%\end{minipage}
\end{figure}

As a next step we consider possible H$_2$ formation via reactions (\ref{eq:h2_one}) and (\ref{eq:last}) on the kite-like and linear isomers of Fe$_2$O$_3$ cluster.
The possible pathways for H$_2$ formation in the case of the most stable ammonia adsorption on the two-coordinated site (I$^{\prime}$ intermediates) of the kite-like Fe$_2$O$_3$ 
isomer are shown in Fig.~\ref{fig:case_1_h2_formation}(a), while the corresponding structures of the optimized equilibrium and 
transition states along the reaction path are illustrated in Fig.~\ref{fig:case_1_h2_formation}(b).

Note that H$_2$ formation can occur after partial decomposition of ammonia in reaction (\ref{eq:h2_one}), 
starting from intermediate (I$^{I}_{1}$3) via the path I$^{\prime}_{1}$3 - T$^{\prime}_{1}$6 - I$^{\prime}_{1}$7 - T$^{\prime}_{1}$7 - I$^{\prime}_{1}$8. 
On the other hand, H$_2$ formation can occur via full decomposition of ammonia in reaction (\ref{eq:last}), through the intermediate (I$^{I}_{1}$4) 
via the path I$^{\prime}_{1}$4 - T$^{\prime}_{1}$4 - I$^{\prime}_{1}$5 - T$^{\prime}_{1}$5 - I$^{\prime}_{1}$6. 
In both cases, the reaction pathways include breaking one O$-$H bond and forming an Fe$-$H bond.
The H$_2$ formation barriers through these intermediates are {89.74 kcal/mol and 92.49 kcal/mol}, respectively. 
{From these results, we conclude that H$_2$ formation on the kite-like Fe$_2$O$_3$ structure is more favorable 
via reaction (\ref{eq:h2_one}), with the NH$^{*}$ intermediate remaining adsorbed on the cluster. 
The H$_2$ formation reaction, starting from (I$^{I}_{1}$4), is the rate-limiting step in molecular hydrogen formation on the kite Fe$_2$O$_3$ cluster.}    

\begin{figure}[H]
%\noindent \begin{minipage}[r]{0.55\textwidth}
\centering
\includegraphics[width=12cm]{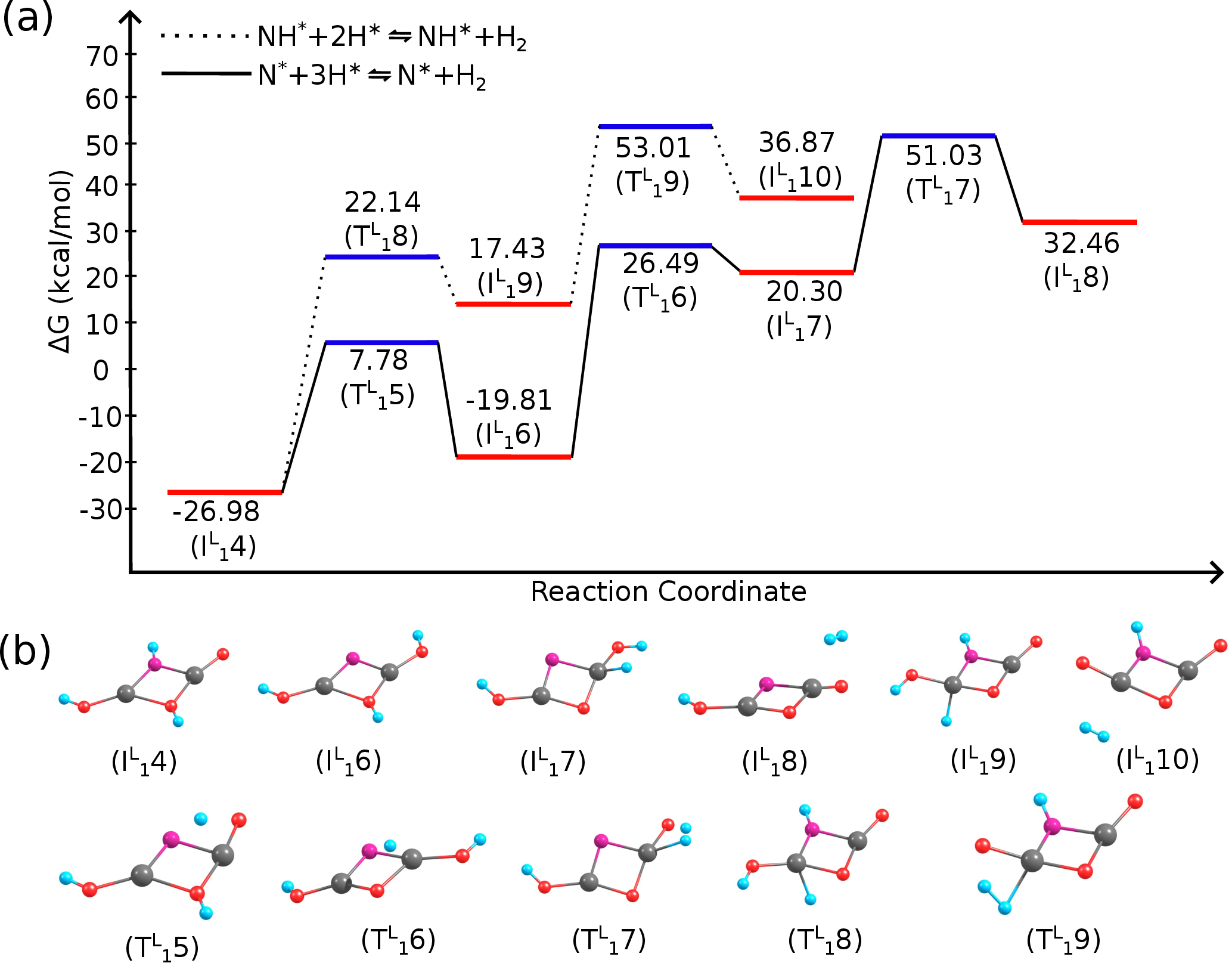}
\caption{(a) The energy profile for H$_2$ formation on the linear isomer of the Fe$_2$O$_3$ cluster at T=298.15 K.
(b) Geometries of the optimized equilibrium and transition states along the reaction path.}
\label{fig:case_2_h2_formation}
%\end{minipage}
%\caption{The energy profile for H$_2$ formation reaction path on (a) kite-like and (b) linear-type of Fe$_2$O$_3$. Geometries of transition state (blue line) and equilibrium geometries (red line) are shown.}
\end{figure}

Similarly, the H$_2$ formation reaction pathways on the linear-type structure of Fe$_2$O$_3$ are shown in Fig.~\ref{fig:case_2_h2_formation}(a), 
while the optimized equilibrium and transition states along the reaction path are illustrated in Fig.~\ref{fig:case_2_h2_formation}(b).
The H$_2$ formation through the NH$^{*}$ intermediate (I$^{L}_{1}$4) via the reaction path 
I$^{L}_{1}$4 - T$^{L}_{1}$8 - I$^{L}_{1}$9 - T$^{L}_{1}$9 - I$^{L}_{1}$10 has an energy barrier of {79.99 kcal/mol. On the other hand 
H$_2$ formation through intermediate (I$^{L}_{1}$6) via reaction path I$^{L}_{1}$6 - T$^{L}_{1}$6 - I$^{L}_{1}$7 - T$^{L}_{1}$7 - I$^{L}_{1}$8 
has an activation energy of 70.84 kcal/mol, 
which is about 10 kcal/mol lower energy than reaction path through intermediate (I$^{L}_{1}$4).}

Overall, on the basis of our calculated reaction pathways for H$_2$ formation show similar pattern for both kite-type and linear-type 
Fe$_2$O$_3$, where H$_2$ formation in reactions (\ref{eq:h2_one}) 
{and (\ref{eq:last}) take place via breaking one of {O$-$H} bond and forming intermediate {Fe$-$H} bond. 
However, from both thermodynamic and kinetic perspectives, H$_2$ formation on the two types of Fe$_2$O$_3$ structures varies. 
Reaction (\ref{eq:h2_one}) is more favorable on the kite-like structure, 
while reaction (\ref{eq:last}) is more favorable on the linear structure. 
This highlights that the rate-limiting step for H$_2$ formation is highly dependent on the catalyst's structure.}

%\hspace{0.2cm}

\smallskip

%######################################################################
%######################################################################
\subsection{\textbf{NH$_3$ decomposition on Fe$_4$O$_6$}}
\label{sec:n2_NH3}

\begin{figure}[H]
\centering
\includegraphics[width=15cm]{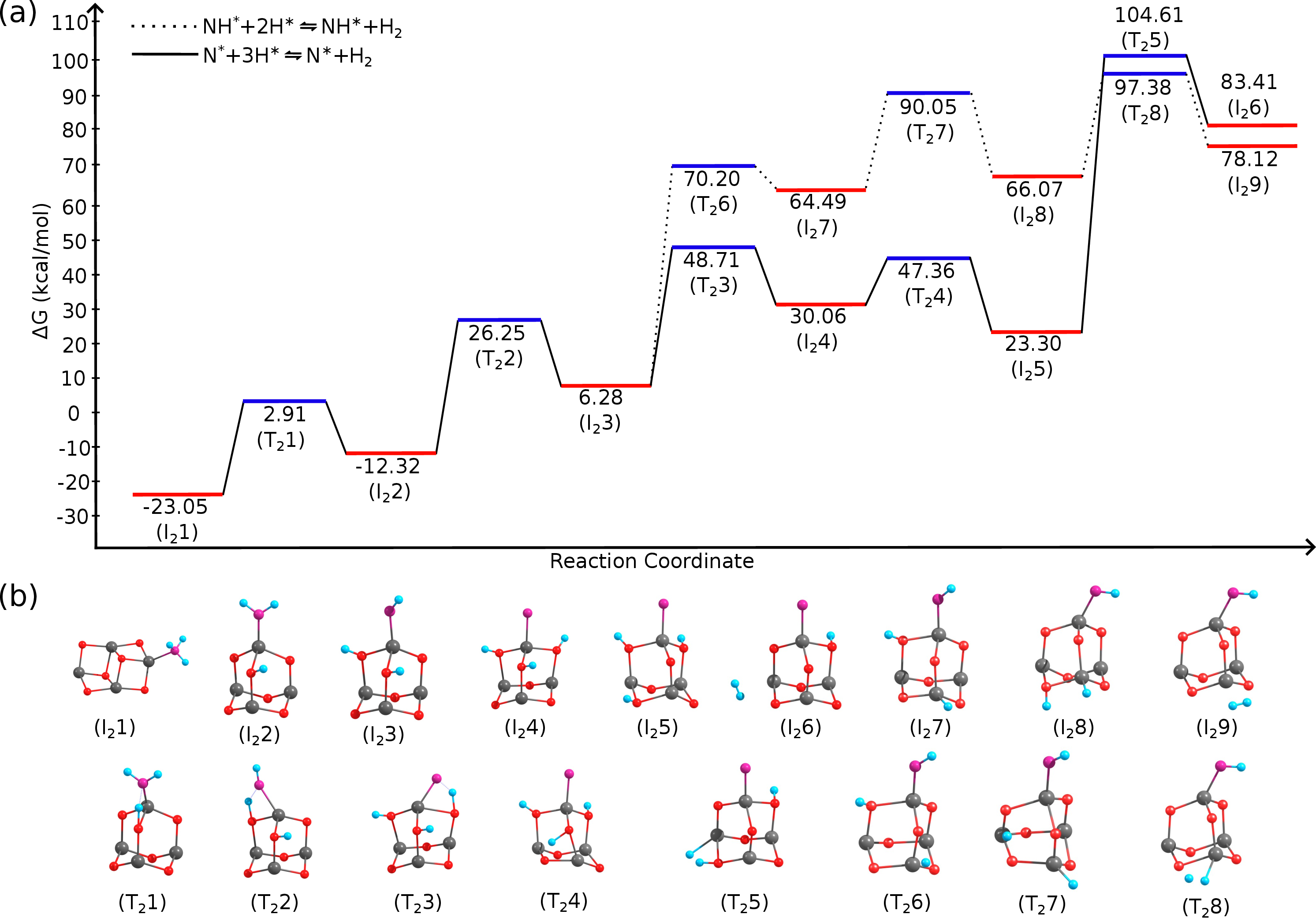}
\caption{(a) The energy profile for NH$^{*}_3$ $\rightarrow$ NH$^{*}_2$ + H$^{*}$ $\rightarrow$ NH$^{*}$ + 2H$^{*}$  $\rightarrow$ N$^{*}$ + 3H$^{*}$  and H$_2$ formation reaction paths on 
the (Fe$_2$O$_3$)$_2$  at T=298.15 K. (b) Geometries of the optimized equilibrium and transition states along the reaction path.}
\label{fig:n2_case1}
\end{figure}

In the following subsection, we discuss the catalytic activity of (Fe$_2$O$_3$)$_2$ towards NH$_3$ dehydrogenation and H$_2$ formation reactions.        
On the basis of adsorption characteristics discussed in \ref{sec:adsorption}, the threefold coordinate Fe$^{3+}$ site of the Fe$_4$O$_6$ cluster is the most stable site for NH$_3$ adsorption. 
Complete reaction pathway for stepwise decomposition of NH$_3$ and formation of H$_2$ reactions on (Fe$_2$O$_3$)$_2$ cluster are depicted in Fig.~\ref{fig:n2_case1}(a), and 
the corresponding intermediate and transition state structures are shown in Fig.~\ref{fig:n2_case1}(b). 
From this point forward, the first dehydrogenation step follows starting 
from the intermediate (I$_2$1) where NH$_3$ molecule interacting with three-coordinated Fe site of (Fe$_2$O$_3$)$_2$ cluster by transferring a hydrogen to its one of neighboring oxygen via 
reaction pathway (I$_2$1 - T$_1$1 - I$_2$2) and reaction barrier of this step is 21.47 kcal/mol which is 5.51 kcal/mol lower energy barrier than                   
first hydrogen transfer on kite-like Fe$_2$O$_3$ cluster. This reaction also involves different isomer of (Fe$_2$O$_3$)$_2$, where decompostion takes place on the second minima 
isomer of (Fe$_2$O$_3$)$_2$ shown in Fig.~{S2}. Relative binding energy of second minima isomer is 2.35 kcal/mol.  
The second dehydrogenation step follows from adsorbate NH$_{2}^{*}$ intermediate (I$_2$2) further dissociate to NH$^{*}$ + 2H$^{*}$ which
dissociated hydrogen atom subsequently transferred to another neighboring oxygen as shown in the reaction path (I$_2$2 - T$_2$2 - I$_2$3). This reaction occurs with 
energy barrier of 38.57 kcal/mol. The ultimate dehydrogenation step is the formation of N$^{*}$ + 3H$^{*}$ where N is bound to the central top Fe$^{3+}$ and all the hydrogen atoms interact with 
three neighboring oxygens. The last dehydrogenation step occurs with the energy barrier 3.86 kcal/mol higher than the energy barrier of the second dehydrogenation step and it is shown 
in the reaction pathway (I$_2$3 - T$_2$3 - I$_2$4). It suggests that dehydrogenation of adsorbate NH$^{*}$ is rate-determining step on (Fe$_2$O$_3$)$_2$ cluster. 
Moreover, from a thermodynamic viewpoint calculated dehydrogenation steps of NH$_3$ on (Fe$_2$O$_3$)$_2$ cluster is endothermic by 6.24, 18.6, and 23.78 kcal/mol.

Considering H$_2$ formation reactions via two reaction pathways. First H$_2$ formation reaction (\ref{eq:h2_one}) occurs 
with partial decomposition of NH$_3$ starting from intermediate (I$_2$3) through (I$_2$9). The first stage through this reaction path starting from (I$_2$3), the transition state (T$_2$6) 
was found that the H atom adsorbed on the Fe atom and formed a {Fe$-$H} bond. In the second stage of the reaction, the transition state (T$_2$7) was the one that splits the adsorbed H atom from the 
adjacent O atom to adsorbed NH$^{*}$. Then, the dissociated H atom was adsorbed in the O atom which is an adjacent atom to the Fe$-$H bond, and at the final stage, 
the dissociative molecular H$_2$ formed, and barrier of this reaction is 91.1 kcal/mol.  

Completed reaction pathway for reaction (\ref{eq:h2_one}) is (I$_2$3 - T$_2$6 - I$_2$7 - T$_2$7 - I$_2$8 - T$_2$8 - I$_2$9).       
The second H$_2$ formation reaction (\ref{eq:last}) is that occurs with fully decomposed NH$_3$ molecule  starting from intermediate (I$_2$4) through intermediate (I$_2$6). 
It is important to note that last dehydrogenation reaction (\ref{eq:fourth}) is the one which has the highest barrier on the (Fe$_2$O$_3$)$_2$ cluster. So dissociative molecular hydrogen formation 
through this reaction path cost an energy as shown in reaction path (I$_2$4 - T$_2$4 - I$_2$5 - T$_2$5 - I$_2$6).    
Overall, as it seen from depicted reaction pathways in Fig.~\ref{fig:n2_case1}, H$_2$ formation reaction is kinetically and energetically costly in reaction 
N$^{*}$ + 3H$^{*}$ $\rightleftharpoons$ N$^{*}$ + H$^{*}$ + H$_{2}$, and it is more favorable via reaction NH$^{*}$ + 2H$^{*}$ $\rightleftharpoons$ NH$^{*}$ + H$_{2}$ which is partial 
decomposition of NH$_3$ on (Fe$_2$O$_3$)$_2$ cluster.    

\smallskip

\bigskip

\subsection{\textbf{NH$_3$ decomposition on Fe$_6$O$_9$}}
\label{sec:n3_NH3}

\begin{figure}[H]
\centering
\includegraphics[width=13cm]{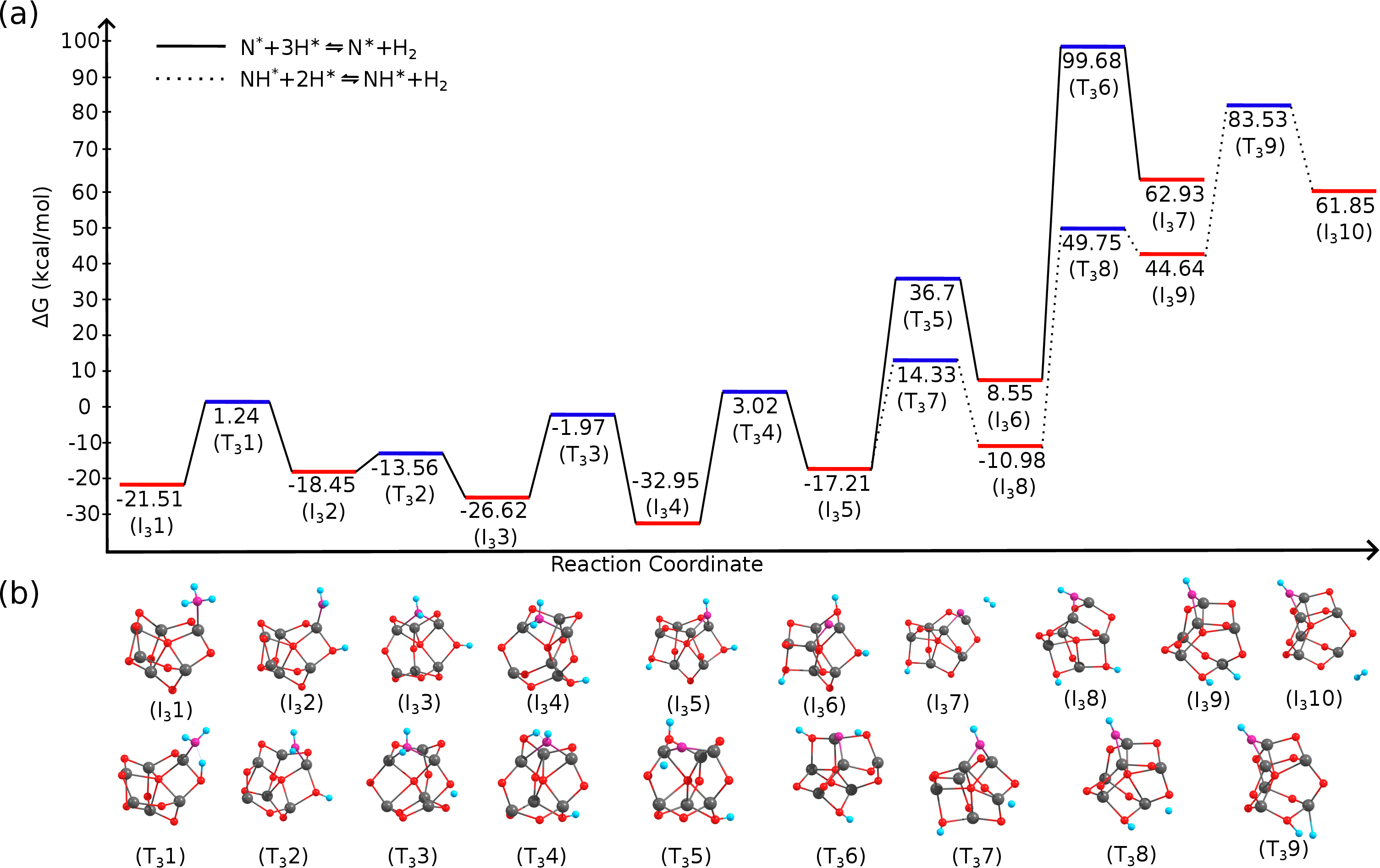}
\caption{(a) The energy profile for NH$^{*}_3$ $\rightarrow$ NH$^{*}_2$ + H$^{*}$ $\rightarrow$ NH$^{*}$ + 2H$^{*}$  $\rightarrow$ N$^{*}$ + 3H$^{*}$ and H$_2$ formation reaction paths on 
the (Fe$_2$O$_3$)$_3$ at T=298.15 K. (b) Geometries of the optimized equilibrium and transition states along the reaction path.}
\label{fig:n3}
\end{figure}

The energy profile for the stepwise dehydrogenation of NH$_3$ on the (Fe$_2$O$_3$)$_3$ cluster is presented 
in Fig.~\ref{fig:n3}(a), while the intermediate and transition state structures along this reaction pathway 
are shown in Fig.~\ref{fig:n3}(b). 
The dissociation of NH$_3$ on the (Fe$_2$O$_3$)$_3$ cluster is more complex compared to smaller Fe(III) oxide structures, 
as NH$_3$ can adsorb at various sites on the (Fe$_2$O$_3$)$_3$ surface.

We identified the most favorable adsorption configuration, I$_3$1, with an adsorption energy of $\Delta G = -21.51$ kcal/mol, 
from which the stepwise decomposition reaction proceeds. 
The first dehydrogenation reaction, as described in (\ref{eq:second}), 
begins with NH$_3^*$ adsorbed on the (Fe$_2$O$_3$)$_3$ cluster as I$_3$1 and proceeds 
through the transition state T$_3$1. 
The energy barrier along this pathway is 22.75 kcal/mol, which is slightly higher than the barrier 
for the first H abstraction from NH$_3$ on the (Fe$_2$O$_3$)$_2$ cluster. 
Although the first dehydrogenation reaction on the (Fe$_2$O$_3$)$_3$ cluster is endothermic, 
we observed that when the NH$_2^*$ species migrates to a bridging position between two Fe atoms (Fe $-$ N $-$ Fe), 
the reaction becomes exothermic by 11.44 kcal/mol, as shown in the reaction 
pathways I$_3$2 $-$ T$_3$2 $-$ I$_3$3 and I$_3$3 $-$ T$_3$3 $-$ I$_3$4.

The second H abstraction involves the further dehydrogenation of NH$_2^*$ into NH$^*$ and H$^*$, 
with an energy barrier of 35.97 kcal/mol along the pathway I$_3$4 $-$ T$_3$4 $-$ I$_3$5. 
This barrier is 10 kcal/mol higher than that of the first dehydrogenation step. 
Additionally, this reaction is endothermic, with a reaction energy of 15.74 kcal/mol.

Similarly, in the third step (\ref{eq:fourth}), the remaining NH$^*$ dissociates into N$^*$ and H$^*$, with an energy 
barrier 17.94 kcal/mol higher than that of the second dissociation step. 
This is the largest barrier encountered in the decomposition of NH$_3$. 
The calculated reaction pathway indicates that this process is endothermic, with a reaction energy of 25.76 kcal/mol.

Lastly, the possible H$_2$ formation reactions (\ref{eq:h2_one} and \ref{eq:last}) on the (Fe$_2$O$_3$)$_3$ cluster were calculated, 
as shown in Fig.~\ref{fig:n3}. The first H$_2$ formation reaction (\ref{eq:h2_one}) 
begins with one adsorbed NH$^*$ and two H$^*$ species on the (Fe$_2$O$_3$)$_3$ cluster. 
The reaction proceeds in a manner similar to that discussed in the previous subsection: 
the adsorbed H$^*$ on oxygen, adjacent to the NH$^*$ adsorbed on Fe, migrates away by forming Fe$-$H bonds 
through transition states T$_3$7 and T$_3$8. 
The overall energy barrier for H$_2$ formation via reaction (\ref{eq:h2_one}) is 100.74 kcal/mol.

The second possible H$_2$ formation pathway starts from fully decomposed NH$_3$ (I$_3$6) 
and proceeds through the transition state T$_3$6. This pathway has a significantly high energy barrier, 
calculated to be 116.89 kcal/mol, as shown in the reaction path I$_3$6 $-$ T$_3$6 $-$ I$_3$7. 
These results suggest that, from both a thermodynamic and kinetic perspective, H$_2$ formation after full 
dehydrogenation of NH$_3$ is less favorable.

\smallskip

\smallskip

\subsection{\textbf{NH$_3$ decomposition on Fe$_8$O$_{12}$}}
\label{sec:n4_NH3}
%----------------------------------------------------------------

\begin{figure}[H]
\centering
\includegraphics[width=13cm]{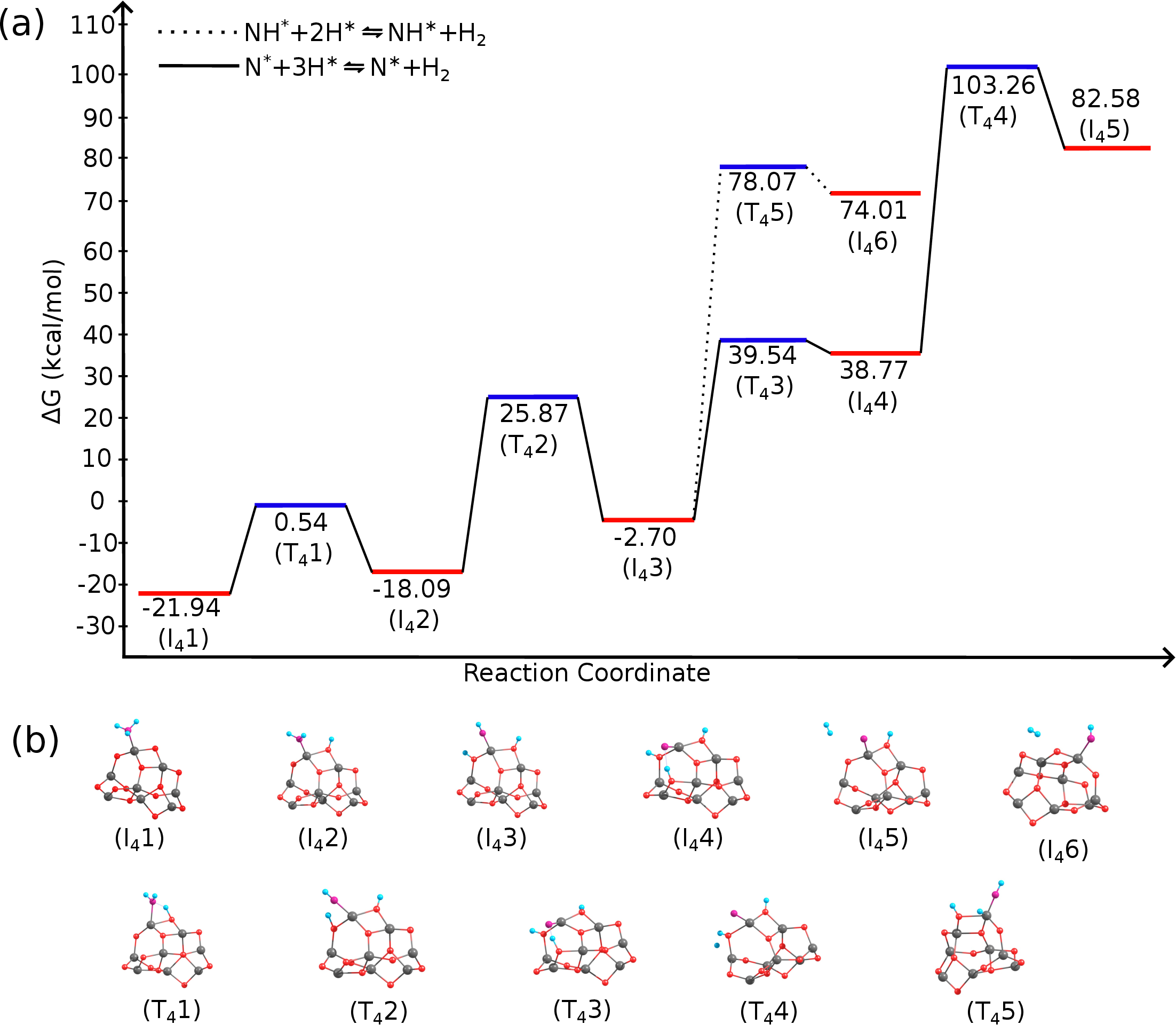}
\caption{(a) The energy profile for NH$^{*}_3$ $\rightarrow$ NH$^{*}_2$ + H$^{*}$ $\rightarrow$ NH$^{*}$ + 2H$^{*}$  $\rightarrow$ N$^{*}$ + 3H$^{*}$  and H$_2$ formation reaction path on 
the (Fe$_2$O$_{3}$)$_4$  at T=298.15 K. (b) Geometries of the optimized equilibrium and transition states along the reaction path.}
\label{fig:n4}
\end{figure}

Finally, the decomposition of NH$_3$ and the H$_2$ formation pathways on the (Fe$_2$O$_3$)$_4$ cluster are 
illustrated in Fig.~\ref{fig:n4}(a), with the intermediate and transition state structures shown in Fig.~\ref{fig:n4}(b). 
As discussed in previous subsections, increasing the number of units $n$ in (Fe$_2$O$_3$)$_n$ increases the number 
of active sites that interact with NH$_3$. However, similar to the reactions on (Fe$_2$O$_3$)$_n$ ($n=2,3$), 
the most stable adsorption site for NH$_3$ on (Fe$_2$O$_3$)$_4$ is a three-coordinated Fe site, 
with an adsorption energy of -21.94 kcal/mol at room temperature, slightly higher than that on (Fe$_2$O$_3$)$_3$.
The dehydrogenation of NH$_3$ begins with the adsorption of NH$_3^*$, as shown in the intermediate state I$_4$1. 
The first dehydrogenation step involves breaking one N$-$H bond and forming an O$-$H bond, 
with an energy barrier of 22.48 kcal/mol, as shown in the reaction pathway I$_4$1 $-$ T$_4$1 $-$ I$_4$2. 
The second dehydrogenation step (\ref{eq:third}) involves the dissociation of NH$_2^*$ + H$^*$ to form NH$^*$ + 2H$^*$, 
proceeding through the transition state T$_4$2. The energy barrier for this step is 43.96 kcal/mol, which is higher 
than the corresponding second dehydrogenation steps on (Fe$_2$O$_3$)$_n$ ($n=1-3$). 
The final dehydrogenation step occurs along the pathway I$_4$3 $-$ T$_4$3 $-$ I$_4$4, with a barrier of 42.24 kcal/mol.
All NH$_3$ dehydrogenation steps on (Fe$_2$O$_3$)$_4$ are endothermic, with reaction 
energies of 3.85 kcal/mol, 15.39 kcal/mol, and 41.47 kcal/mol, respectively.

The final reaction pathway on the (Fe$_2$O$_3$)$_4$ cluster involves H$_2$ 
formation from both partially and fully decomposed NH$_3$, as described in (\ref{eq:h2_one}) and (\ref{eq:last}). 
As observed for all sizes of (Fe$_2$O$_3$)$_n$ clusters, H$_2$ formation is energetically more 
favorable after the partial decomposition of NH$_3$ in reaction (\ref{eq:h2_one}) compared to the fully 
decomposed pathway (\ref{eq:last}). However, this pathway also presents the highest energy barrier on this cluster.

\smallskip

\smallskip

%-------------------------------------------------------------
%============================================================
%=========Comparison and conclusion==========================
%============================================================

\section{Comparison and conclusion}
\label{sec:comp}

Our results, illustrated in Fig.~\ref{fig:case_1}, Fig.~\ref{fig:case_2}, Fig.~\ref{fig:n2_case1}, Fig.~\ref{fig:n3}, 
and Fig.~\ref{fig:n4}, indicate that NH$_3$ dehydrogenation can be a thermodynamically favorable reaction 
on (Fe$_2$O$_3$)$_n$ ($n=1-4$) clusters. However, the favorability depends on the size and geometry of the cluster, 
as well as the specific reaction steps described in (\ref{eq:second}) $-$ (\ref{eq:last}).

To compare the activity of various sizes and structures of (Fe$_2$O$_3$)$_n$ ($n=1-4$), 
we have calculated the change in Gibbs free energy ($\Delta G$) as a function of temperature at 1 bar pressure, 
as shown in Fig. S6. Across all reactions studied, we observed that $\Delta G$ 
increases with temperature. This suggests that NH$_3$ dehydrogenation on (Fe$_2$O$_3$)$_n$ ($n=2,4$) can be energetically 
favorable at moderate temperatures, depending on the specific reaction step. However, as the temperature rises 
beyond a certain threshold, the reaction becomes unfavorable.

For example, as shown in Fig. S6 (a), (b), and (c), all dehydrogenation reactions 
on (Fe$_2$O$_3$)$_n$ ($n=1$) are energetically favorable within the temperature range of 0$-$1000 K. 
In contrast, on (Fe$_2$O$_3$)$_n$ ($n=2,4$), only the last dehydrogenation step is limiting. 
Since the $\Delta G$ of the third dehydrogenation reaction is already greater than zero at 0 K, this step is not 
favorable at any temperature.
Another larger cluster considered in this study, (Fe$_2$O$_3$)$_n$ ($n=3$), exhibits better stability of the reaction intermedeates during 
the second dehydrogenation step, remaining favorable up to 800 K. On the other hand, 
the second dehydrogenation reaction on (Fe$_2$O$_3$)$_n$ ($n=4$) is favorable only up to 400 K.
The most endothermic dehydrogenation reaction on this cluster is the step NH$^*$ $\rightleftharpoons$ N$^*$ + 3H$^*$. 
The first and second dehydrogenation steps are favorable up to 1100 K and 700 K, respectively.

Moreover, we observed the variation of $\Delta G$ with temperature for the H$_2$ formation 
reaction on (Fe$_2$O$_3$)$_n$ ($n=1-4$). Our results indicate that the formation of molecular hydrogen is not 
thermodynamically favorable at any temperature. However, temperature is not the only factor 
determining whether the reaction occurs. If sufficient energy is available to overcome the activation barrier, 
the reaction can still proceed.

%%%%%%%%%%%%%%%%%%%%%%%%%%%%%%%%%%%%%%%%%%%%%%%%%%%%%%%%%%%%%%%%%%%%%%
\iffalse 

\smallskip

\begin{figure}[H]
\noindent \begin{minipage}[r]{0.45\textwidth}
\centering
\includegraphics[width=7cm]{figures/delta_G_reaction_2_NH3_to_NH2_in_K.eps}
\subcaption{NH$_{3}^{*}$ + $^{*} \rightleftharpoons$ NH$_{2}^{*}$ + H$^{**}$ }
\label{fig:step_2}
\end{minipage}
\hspace{0.2cm}
\noindent \begin{minipage}[r]{0.55\textwidth}
\centering 
\includegraphics[width=7.0cm]{figures/delta_G_reaction_3_NH2_to_NH_in_K.eps}
\subcaption{NH$_{2}^{*}$ + $^{*} \rightleftharpoons$ NH$^{*}$ + 2H$^{**}$ } 
\label{fig:step_3}
\end{minipage} 
\vspace{1.0cm}
\noindent \begin{minipage}[r]{0.45\textwidth}
\centering
\includegraphics[width=7.0cm]{figures/delta_G_reaction_4_NH_to_N_in_K.eps}
\subcaption{NH$^{*}$ + $^{*} \rightleftharpoons$ N$^{*}$ + 3H$^{**}$}
\label{fig:step_4}
\end{minipage}
\hspace{0.2cm}
\noindent \begin{minipage}[r]{0.55\textwidth}
\centering
\includegraphics[width=7.0cm]{figures/delta_G_reaction_5_H2_formation_in_K.eps}
\subcaption{NH$^{*}$ + $^{*} \rightleftharpoons $ NH$^{*}$ + H$_{2}$ }
\label{fig:step_5}
\end{minipage}
\caption{Variation of Gibbs free energy ($\Delta G$) with temperature for each dehydrogenation step of NH$_3$ on 
(Fe$_2$O$_3$)$_n$ ($n=1-4$) clusters, along with the H$_2$ formation reaction (\ref{eq:h2_one}).}
\label{fig:reactions_2345}
\end{figure}

\fi
%%%%%%%%%%%%%%%%%%%%%%%%%%%%%%%%%%%%%%%%%%%%%%%%%%%%%%%%%%%%%%%%%%%%%%%%%%%%%%
\smallskip 

The effective production of molecular hydrogen from ammonia is determined by the stepwise 
dehydrogenation of adsorbed ammonia on the catalyst. Catalytic reaction mechanisms are analyzed by 
identifying the rate-determining step in the dehydrogenation of NH$_3$, which corresponds to the step requiring 
the highest energy to activate the N$-$H bond. However, it is important to note that in catalysis, 
the overall energy barrier is more significant than the barrier for any single intermediate reaction step.

\begin{figure}[H]
\centering
\includegraphics[width=13cm]{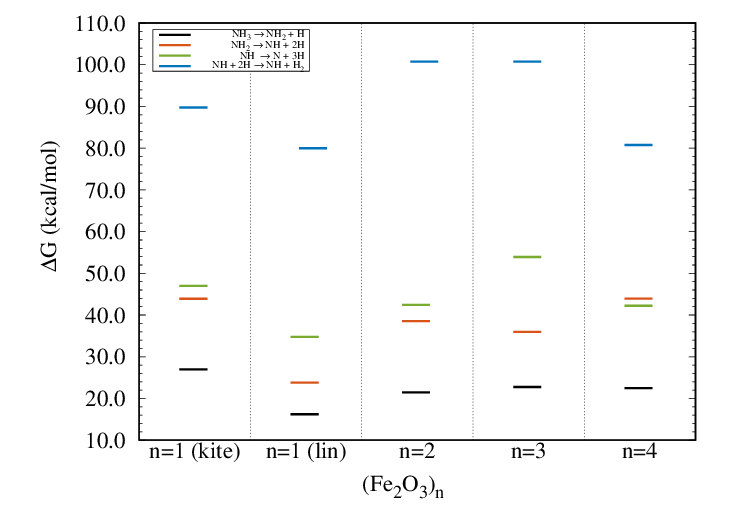}
\caption{Reaction barrier ($\Delta$G$^{\ddag}$) for NH$_3$ dehydrogenation and H$_2$ formation reactions on (Fe$_2$O$_{3}$)$_n$ ($n=1-4$) clusters.}
\label{fig:barrier}
\end{figure}

Several studies have reported different rate-determining steps depending on the type of 
catalyst used\cite{kulkarni2023elucidating}. Xiuyuan Lu et al. found that the rate-determining step 
in NH$_3$ decomposition on different phases of Ru surface catalysts is the formation of molecular nitrogen\cite{lu2021kinetic}. 
In contrast, studies by Xilin Zhang et al.\cite{zhang2015adsorption} on ammonia decomposition on small iron clusters 
showed that the rate-determining step on single Fe and Fe$_3$ is the NH $\rightarrow$ N + H step, 
whereas for Fe$_2$ and Fe$_4$, the rate-determining step is the NH$_2$ $\rightarrow$ NH + H step.
Similarly, a detailed comparison of the energy barriers for each elementary step in NH$_3$ decomposition and H$_2$ 
formation on different sizes and shapes of (Fe$_2$O$_3$)$_n$ ($n=1-4$) is shown in Fig.~\ref{fig:barrier}. 
Based on the results from our calculations, the rate-determining step in ammonia decomposition and H$_2$ 
formation varies with the size of the (Fe$_2$O$_3$)$_n$ ($n=1-4$) oxide clusters.
In general, the final step of H$_2$ formation represents the highest energy barrier on all 
(Fe$_2$O$_3$)$_n$ ($n=1-4$) clusters. However, the analysis of NH$_3$ decomposition shows 
that the NH $\rightarrow$ N + H step is typically the rate-determining step, except in the case of (Fe$_2$O$_3$)$_4$, 
where the rate-determining step is the second H dissociation step.
Furthermore, the first dehydrogenation step exhibits an energy barrier that is nearly identical across all clusters, 
with the process being exothermic for clusters $n=1$ and $n=3$, and endothermic for clusters $n=2$ and $n=4$. 
For the second dehydrogenation step, (Fe$_2$O$_3$)$_3$ demonstrates significantly higher activity compared 
to the other cluster sizes. 
It is also important to note that $n=1$ (linear) is the only special configuration of Fe$_2$O$_3$ 
containing two terminal O$^{-2}$ ions, unlike the other types of Fe$_2$O$_3$, which may promote a potentially 
high activity for NH$_3$ dehydrogenation and molecular hydrogen formation. 
Overall, the lowest energy barrier observed for H$_2$ formation is associated with the largest cluster considered in this study.

\smallskip 

\smallskip

In this research, various structures of (Fe$_2$O$_3$)$_n$ ($n=1-4$) were obtained using the SC-AFIR method, 
and we investigated the ammonia decomposition and molecular hydrogen formation reaction pathways on the most stable 
isomers of (Fe$_2$O$_3$)$_n$ ($n=1-4$) clusters. This analysis employed the SC-AFIR and DS-AFIR methods 
within the Global Reaction Route Mapping (GRRM) strategy, utilizing the B3LYP exchange-correlation functional in Kohn-Sham DFT.

The results indicate that the catalytic activity in ammonia decomposition varies depending on the 
size and shape of the high-spin iron trioxides. The adsorption analysis reveals that the NH$_3$ molecule preferentially 
adsorbs at two-coordinated Fe sites in $n=1$, and at three-coordinated Fe sites in $n=2-4$ clusters. 
Furthermore, the adsorption energy tends to decrease from $n=1$ to $n=3$ of the (Fe$_2$O$_3$)$_n$ clusters, 
then slightly increases for the (Fe$_2$O$_3$)$_4$ cluster.

From a thermodynamic perspective, the adsorption of the NH$_3$ molecule on (Fe$_2$O$_3$)$_1$ is 
favorable across the entire temperature range of 0 K to 1200 K. In contrast, for the larger 
clusters (Fe$_2$O$_3$)$_n$ ($n=2, 4$), ammonia {adsorption} becomes energetically unfavorable 
at temperatures of 1107 K, 961 K, and 1000 K for $n=2, 3$, and $4$, respectively.

A comparison of the rate-determining steps in the ammonia dehydrogenation reaction reveals a dependency on the 
size of the iron trioxide clusters. Thus, the reaction step NH$^{*} \rightarrow$ N$^{*} + H^{*}$ is 
the rate-determining step for the smaller iron trioxide clusters (Fe$_2$O$_3$)$_n$ ($n=1-3$). 
In contrast, the reaction step NH$^{*}_{2} \rightarrow$ NH$^{*} + H^{*}$ is identified as the rate-determining 
step for the (Fe$_2$O$_3$)$_n$ ($n=4$) cluster. 
Additionally, we observed that the energy barrier for molecular hydrogen formation increases with 
the size of the clusters (Fe$_2$O$_3$)$_n$ ($n=1-3$) but then experiences a drastic decrease for the (Fe$_2$O$_3$)$_4$ cluster.

We have investigated the catalytic activity of high-spin (Fe$_2$O$_3$)$_n$ ($n=1-4$) clusters 
for decomposition of NH$_3$. We believe that the results are valuable for designing iron trioxide-based nanosized 
catalysts by regulating the size of the (Fe$_2$O$_3$)$_n$ clusters to enhance H$_2$ production from the catalytic 
decomposition of ammonia.

\begin{acknowledgments}
This work was partly supported by MEXT Program: Data Creation and Utilization-Type Material Research and 
Development Project Grant Number JPMXP1122712807, and {partially supported by NAWA "STE(E)R-ING towards International Doctoral School"} 
Calculations were performed using computational resources of the Institute for Solid State Physics, the University of Tokyo, Japan, 
and the Research Center for Computational Science, Okazaki, Japan (Project: 23-IMS-C016).
S.I. is grateful to the MANABIYA system of the Institute for Chemical Reaction Design and Discovery (ICReDD) 
of Hokkaido University, which was established by the World Premier International Research Initiative (WPI), MEXT, 
Japan, to support the learning of the GRRM program techniques for DFT calculations.  
\end{acknowledgments}

\nocite{*}
\bibliography{nh3_decomposition_ref}% Produces the bibliography via BibTeX.

\newpage
\includepdf[pages=1]{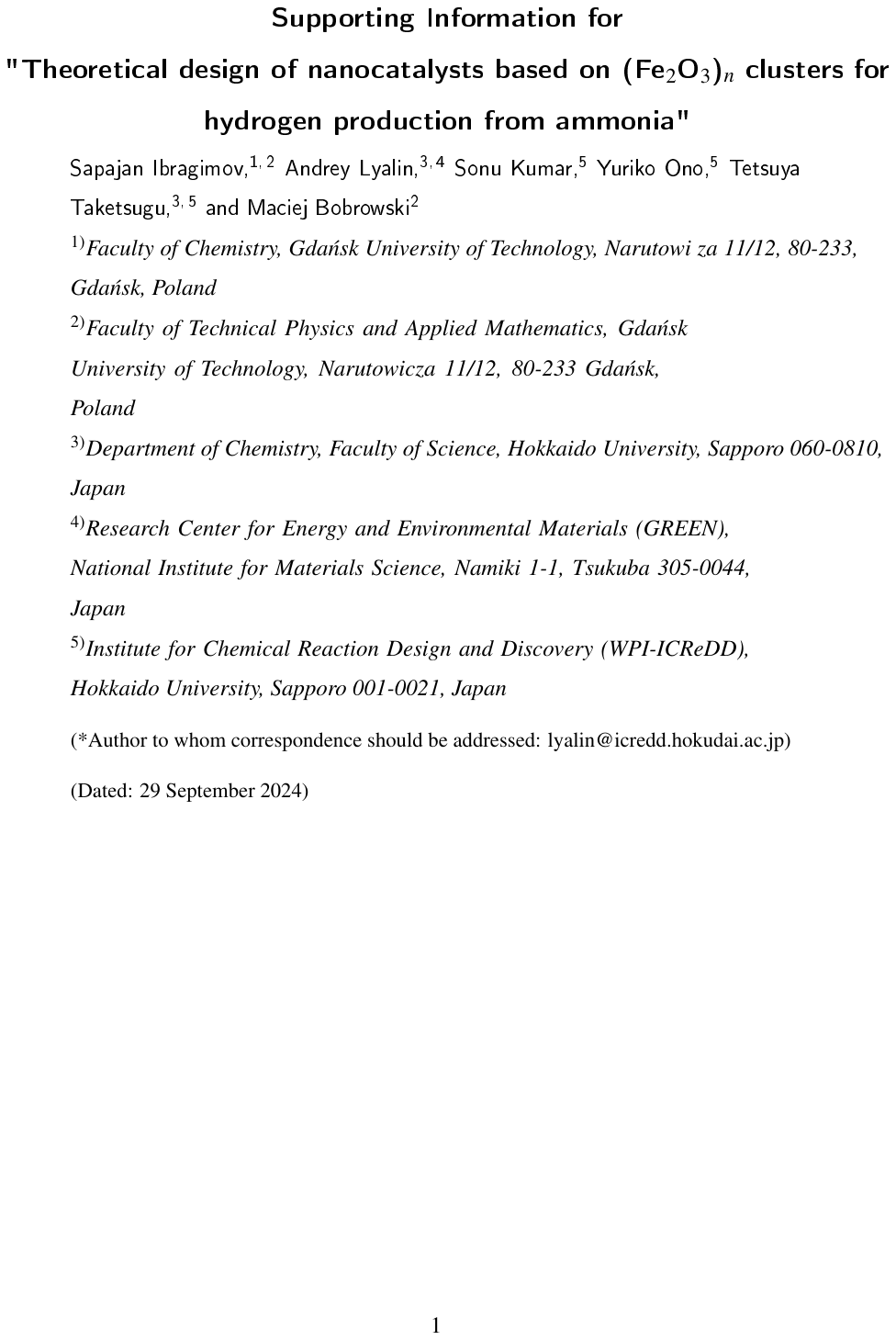}
\newpage
\includepdf[pages=2]{SINH3Fe2O3n.pdf}
\newpage
\includepdf[pages=3]{SINH3Fe2O3n.pdf}
\newpage
\includepdf[pages=4]{SINH3Fe2O3n.pdf}

\end{document}